\documentclass[a4paper,11pt]{article}

\usepackage{a4}
\usepackage{jheppub}
\usepackage{graphics}
\usepackage{amssymb,amsmath}
\usepackage{color}
\usepackage{axodraw}
\usepackage{wasysym}
\usepackage[tight]{subfigure}
\usepackage{booktabs}

\newcommand{\mneu}[1]{m_{\tilde{\chi}^0_{#1}}}

\newcommand{\fb}{\ \mathrm{fb}}
\newcommand{\gev}{\ \mathrm{GeV}}

\newcommand{\met}{$E_T^\mathrm{miss}$}

\preprint{DESY-12-116}
\title{Constraining compressed supersymmetry using leptonic signatures}
\author{Krzysztof Rolbiecki}
\author{and Kazuki Sakurai}
\affiliation{DESY, Notkestrasse 85, D-22607 Hamburg, Germany}
\emailAdd{krzysztof.rolbiecki@desy.de}
\emailAdd{kazuki.sakurai@desy.de}
\abstract{We study the impact of the multi-lepton searches at the LHC on supersymmetric models with compressed mass spectra. For such models the
acceptances of the usual search strategies are significantly reduced due to requirement of large effective mass and \met. On the other hand, 
lepton searches do have much lower thresholds for \met~and $p_T$ of the final state objects. Therefore, if a model with a compressed
mass spectrum allows for multi-lepton final states, one could derive constraints using multi-lepton searches.
For a class of simplified models we study the exclusion limits using ATLAS multi-lepton search analyses
for the final states containing 2--4 electrons or muons
with a total integrated luminosity of 1--2~$\fb^{-1}$ at $\sqrt{s}=7$~TeV. 
We also modify those analyses by imposing additional cuts, so that their sensitivity to compressed supersymmetric models increase.
Using the original and modified analyses, 
we show that the exclusion limits can be competitive with jet plus \met  ~searches,
providing exclusion limits up to gluino masses of 1~TeV.   
We also analyse the efficiencies for several classes of events coming from different intermediate state particles.
This allows us to assess exclusion limits in similar class of models with different cross sections and branching ratios
without requiring a Monte Carlo simulation.
}

\begin{document}
\maketitle
\flushbottom

\section{Introduction\label{intro}}

Uncovering the mechanism of the electroweak symmetry breaking is one of the primary goals of the Large Hadron Collider (LHC). In the Standard Model (SM), the size of the electro-weak scale is determined by the Higgs mass parameter, which receives a large radiative corrections of the order of a cut off scale. Very precise tuning is required on the bare Higgs mass parameter so that it cancels the radiative correction down to the correct size of the weak scale, ${\cal O}(100)$~GeV. This fine tuning is considered to be unnatural and new physics which naturally explains the hierarchy between the weak scale and the cut off scale is anticipated.   
 
Supersymmetry (SUSY)~\cite{susy1,susy2,susy3} is one of the most promising candidates of such new physics models. It reduces the radiative correction to the Higgs mass parameter up to the SUSY mass scale and the fine tuning is not required as long as the mass scale of the SUSY particles is close to the weak scale.
Therefore, it is expected that SUSY should be discovered at the LHC, if it solves the hierarchy problem.
 
One of the difficulties of the SUSY searches is its large volume of the parameter space. Even the Minimal SUSY extension of the Standard Model (MSSM) contains more than 100 free parameters. The number of free parameters can be reduced to ${\cal O}(20)$ if one adopts an ansatz deduced from flavour constraints~\cite{pMSSM}, but a brute force scan of the full parameter space is still not feasible.  

One way to avoid this problem is to introduce further simplifications.
A typical example is the constrained MSSM (CMSSM)~\cite{cmssm1,cmssm2,cmssm3}.   
It assumes a common mass $m_0$ ($m_{1/2}$) for all sfermions (gauginos) at the grand unified theory scale.
The low energy sparticle spectrum is then calculated through the renormalisation group evolution, 
which generally leads to large mass hierarchy between coloured and non-coloured SUSY particles.    

The characteristic collider signature of the CMSSM is large missing energy plus multiple high-$p_T$ jets, which originate from 
heavy coloured particles decays to a lighter and neutral lightest SUSY particle (LSP).
The experimental collaborations have eagerly looked at this search channel 
and their negative results have put a stringent constraints on the coloured SUSY particles masses~\cite{atlas_jet_etmiss,cms_jet_etmiss}, which are typically greater than 1~TeV.
This constraint has already revealed that the CMSSM has to possess some level of fine tuning,\footnote{
If the Higgs boson is discovered with its mass around 125~GeV, it will also result in a serious problem in realising such a relatively heavy Higgs mass 
without introducing a precise parameter tuning in the framework of the MSSM. 
This problem may be relaxed by introducing additional chiral superfields or an extra U(1) gauge symmetry;
see e.g.\ refs.~\cite{extended_higgs,nmssm}.} and indicates that SUSY searches based on the CMSSM may not be the right direction to pursue a sign of supersymmetry at the LHC.

Recently, non-universal sfermion mass models (with heavy first two generation squarks plus light third generation squarks)
have attracted attention~\cite{nusm1,nusm2,nusm3,nusm4,nusm5,nusm6}. 
In this class of models, the mass bound on gluino and third generation squarks are not as severe as in the CMSSM,
because of the smaller cross sections of the gluino and third generation squarks pair production and lower $p_T$ distributions of their decay products.
The tension between naturalness and direct SUSY search constraints is, thus, alleviated in this scenario.

Another interesting possibility is a scenario called ``compressed supersymmetry"~\cite{comp1,comp2,comp3,comp4}, in which 
the mass hierarchy between coloured SUSY particles and the LSP is compressed.
In this scenario, $p_T$ of the SUSY decay products can be very small, and that makes it difficult 
to discriminate the signal from the SM background 
when applying a standard search channel of the multiple high $p_T$ jets plus large missing energy.
The dependence of the efficiency in this search channel on the compression parameter, which is essentially the gluino and LSP mass 
ratio,\footnote{The precise definition of the compression parameter $c$, is $M_1 = (\frac{1+5c}{6}) M_{\tilde g}$;  
see refs.~\cite{comp_lhc1,comp_lhc2} for more details.} 
has been studied in detail in refs.~\cite{comp_lhc1,comp_lhc2}.       

However, the multiple high-$p_T$ jets plus large missing energy search channel is obviously not optimised for the compressed SUSY scenario and it would be interesting to see how stringent constraints can be obtained from other search analyses. The lepton searches typically have lower $p_T$ thresholds for signal particles and missing transverse energy. QCD background can also be significantly suppressed. In this paper, we concentrate on the ATLAS multi-lepton search analyses~\cite{Atlas_2lep, Atlas_3lep, Atlas_4lep}. Recently several studies have investigated constraints on a particular SUSY model by reinterpreting the experimental search results~\cite{amsb,gmsb1,gmsb2,pmssm_lhc,so10}. In this paper, we do not only reinterpret the results of experimental search analyses 
but also modify those analyses to increase the sensitivity to the compressed SUSY scenario and actually extract a better constraint than the one obtained from the original analyses.

The resulting constraints are, however, necessarily dependent on the details of the models. The extracted bounds are not directly applicable to the other models with different cross sections and branching ratios. To overcome this limitation, we decompose the events into several classes each of which has different intermediate state particles, and evaluate the signal selection efficiency for each class of events. This allows us to reconstruct an approximate visible cross section in a similar class of models without doing Monte Carlo (MC) simulation and to obtain exclusion limits of that model 
by comparing it with the reported 95\% CL upper bound.

The paper is organised as follows. In the next section, we define a simplified model, which provides lepton-rich signatures and allows us to investigate signal selection efficiencies of multi-lepton search analyses effectively. Our simulation setup is also explained. The details of the multi-lepton searches are described in section~\ref{sec:3}. In section~\ref{sec:dist}, we study the lepton $p_T$ and missing energy in the compressed SUSY scenario and introduce the possible modifications of the ATLAS multi-lepton analyses to increase the sensitivity. Our main result, namely the 95\% upper limit on the coloured particles masses as a function of the gluino and LSP mass difference, is presented in section~\ref{sec:5}. In section~\ref{sec:6}, we provide a recipe to evaluate an approximate visible cross section, which allows us to calculate exclusion limits in another model without doing MC simulation. Section~\ref{summary} is devoted to summary and conclusion.

\section{Analysis setup \label{sec:2}}

For our study, it is useful to define a simplified model, in which the number of parameters is reasonably small,
and leptons are frequently produced from SUSY cascade decays.
To this end, we define the following model with only two free parameters.
One of the free parameters is the gluino mass, $m_{\tilde g}$.
We assume that first two generations of squarks and gluino are mass degenerate, $m_{\tilde q} = m_{\tilde g}$.
This parameter mainly determines the signal cross section.
The other parameter is a mass difference between the gluino and the LSP, $\Delta m = m_{\tilde g} - m_{\tilde \chi_1^0}$,
which affects the signal efficiency.
For simplicity, we explicitly decouple all the third generation sfermions, $\tilde t_i$, $\tilde b_i$ and $\tilde \tau_i$ ($i=1,2$), as well as the higgsino states. 
In this case, the lighter neutralinos and charginos are purely composed of the gaugino states,
$\tilde \chi^0_1 \simeq \tilde B$, $\tilde \chi_2^0 \simeq \tilde W^0$, $\tilde \chi_1^{\pm} \simeq \tilde W^{\pm}$, with
$m_{\tilde B} = m_{\tilde \chi_1^0}$ and $m_{\tilde W} = m_{\tilde \chi_2^0} = m_{\tilde \chi_1^{\pm}}$.

Leptons can be produced from decays of the wino states:
\begin{equation}
\tilde \chi^0_2 \to \tilde \ell^{\pm} \ell^{\mp} \to \ell^{\pm} \ell^{\mp} \tilde \chi_1^0\;, \qquad \qquad
\tilde \chi^\pm_1 \to \tilde \ell^{\pm} \nu_\ell (\tilde \nu_\ell \ell^{\pm})  \to \ell^{\pm} \nu_\ell \tilde \chi_1^0\;,
\label{eq:decay}
\end{equation} 
To make these decays possible, we place $m_{\tilde W}$ in between the gluino mass and the LSP mass, $m_{\tilde W} = m_{\tilde B} + \Delta m/2$,
and the slepton mass in between the wino mass and the bino mass, $m_{\tilde \ell} = m_{\tilde B} + \Delta m/4$.
The sneutrino mass is taken to be equal to the slepton mass for simplicity. 
We also assume that the left handed squarks, $\tilde q_L$, exclusively decay to the wino states.
The branching ratios of the relevant SUSY particles are listed in table~\ref{tab:br}.

\begin{table}[t]
\renewcommand{\arraystretch}{1.5}
\begin{center}
\begin{tabular}{c|c|c|c|c|c|c|c|c|c|c} 
\toprule
mother particle & $\tilde \ell^{\pm}$   &  $ \tilde \chi^0_2 $  & \multicolumn{2}{|c|}{$\tilde \chi_1^{\pm}$}  & $\tilde q_R$  &  \multicolumn{2}{|c|}{$\tilde q_L$} & \multicolumn{3}{|c}{$\tilde g$}      \\  \hline
decay mode & $\tilde \chi_1^0 \ell^{\pm}$   &  $ \tilde \ell^{\pm} \ell^{\mp}$  & $\tilde \ell^{\pm} \nu_\ell$  & $\tilde \nu_\ell  \ell^{\pm}$  & $\tilde \chi_1^0 q$   & $\tilde \chi_2^0 q$ & $\tilde \chi_1^\pm q$ & $\tilde \chi_1^0 qq$ & $\tilde \chi_2^0 qq$ & $\tilde \chi_1^\pm qq$          \\  \hline 
$BR (\%)$ & 100   &  100  & 50  & 50   & 100  & 33  & 67   &  50.5 & 16.5 & 33.0          \\ 
\bottomrule
\end{tabular}
\end{center}
\renewcommand{\arraystretch}{1.0}
\caption{Branching ratios of the relevant SUSY particles in the simplified model.
 \label{tab:br}}
\end{table}

In our analysis, SUSY events are generated using \texttt{Herwig++~2.5.2}~\cite{Bahr:2008pv,Gieseke:2011na,Gigg:2007cr}
with $\sqrt{s} = 7$~TeV.
The detector response is simulated by \texttt{Delphes~2.0.2}~\cite{Ovyn:2009tx} using the ATLAS detector card.
Following the ATLAS analyses, we use the anti-$k_T$ jet algorithm with a radius parameter of 0.4.
Only jets with $p_T > 20$~GeV and $|\eta| < 2.8$ are considered.
For the lepton isolation, we require the sum of the transverse energy deposited within a cone of $\Delta R < 0.2$ around the
lepton candidate (excluding the lepton candidate itself) to be less than 1.8~GeV, where $\Delta R \equiv \sqrt{(\Delta \eta)^2 + (\Delta \phi)^2}$. 
The electron (muon) must have $p_T > 10$~GeV and $|\eta| < 2.47$ (2.4).
If a jet and an electron are both identified within $\Delta R < 0.2$ of each other, the jet is discarded. 
Furthermore, identified electrons or muons are discarded if the separation from the closest remaining jet is $\Delta R < 0.4$. 
Electrons and muons separated by $\Delta R < 0.1$ are both discarded.

In our compressed SUSY scenario, sleptons and electroweak gauginos are as heavy as gluino and squarks and
their production cross section is negligible in our study.
We, therefore, exclusively generate coloured SUSY particles production events.
The lepton sources are then identified as:
\begin{align}
&\tilde q_L \to \tilde \chi_2^0 q \to \tilde \ell^{\pm} \ell^{\mp} q \to \ell^{\pm} \ell^{\mp} q \tilde \chi_1^0  & BR=33\% \;, \nonumber \\
&\tilde q_L \to \tilde \chi_1^{\pm} q \to \tilde \ell^{\pm} \nu_\ell q \,(\tilde \nu_\ell \ell^{\pm} q) \to \ell^{\pm} \nu_\ell q \tilde \chi_1^0  & BR=67\%\;, \nonumber \\
&\tilde g \to \tilde \chi_2^0 qq \to \tilde \ell^{\pm} \ell^{\mp} qq \to \ell^{\pm} \ell^{\mp} qq \tilde \chi_1^0  & BR \simeq 16\% \;,\nonumber \\
&\tilde g \to \tilde \chi_1^{\pm} qq \to \tilde \ell^{\pm} \nu_\ell qq \,(\tilde \nu_\ell \ell^{\pm} qq)  \to \ell^{\pm} \nu_\ell qq \tilde \chi_1^0   & BR \simeq 33\%\;. 
\label{eq:br}
\end{align}

\section{ATLAS multi-lepton searches \label{sec:3}}

In order to constrain the compressed SUSY scenario, in this paper we focus on three ATLAS analyses using multi-lepton signatures,\footnote{
CMS multi-lepton searches~\cite{cms_ss} may also be relevant to the compressed SUSY scenario.
However, because of a lower $p_T$ thresholds used there (with $p_T>10$~GeV and $p_T>5$~GeV for electrons and muons, respectively)
those searches require different implementation of isolation algorithm, and therefore we leave it for future studies.
}
which are categorized by
the number of leptons: two~\cite{Atlas_2lep}, three~\cite{Atlas_3lep} 
and four or more leptons~\cite{Atlas_4lep}.
These analyses do not use any high-$p_T$ jet requirements and the cut on the missing transverse energy is relatively low.    
Data samples are collected with a single muon or electron trigger: at least one muon with $p_T > 20$~GeV or electron with $p_T>25$~GeV is required 
in the all analyses.

For the di-lepton analysis, we use two signal regions defined in ref.\,\cite{Atlas_2lep}: 
``same sign inclusive'' (2SS) and ``opposite sign inclusive'' (2OS) signal 
regions.\footnote{Those signal regions are labelled as ``SS-inc'' and ``OS-inc'' in ref.~\cite{Atlas_2lep}. 
In what follows we will refer to them as ``2SS'' and ``2OS''.}
In both signal regions, events must have exactly two isolated leptons of invariant mass greater than 12\,GeV.
The missing transverse energy, $E_T^{\rm miss}$, must be greater than 250\,GeV in 2OS signal region, whilst $E_T^{\rm miss} > 100$\,GeV in 2SS signal region. 
These signal regions should be most sensitive to the events with $\tilde \chi_1^\pm\tilde \chi_1^\pm$ intermediate state since those produce exactly two charged leptons 
with relatively large branching ratios; see eq.~\eqref{eq:br}.
The other class of events can also contribute to the signal regions after taking into account the lepton identification/isolation efficiency.   

The tri-lepton analysis (3LEP) requires exactly three isolated leptons, of which at least one pair must be Same Flavour and Opposite Sign (SFOS).
If a SFOS lepton pair has invariant mass less than 20\,GeV or within the $Z$-mass window, 81\,GeV$<m_{\ell\ell}<$101\,GeV, the event is discarded.
The event cannot have a $b$-jet and $E_T^{\rm miss}$ is required to be greater than 50\,GeV.   
This signal region should provide a good sensitivity to the events with $\tilde \chi_1^\pm\tilde \chi_2^0$ intermediate state since
it produces exactly three charged leptons. 

In the four-lepton analysis (4LEP), an event must contain four or more isolated leptons.
Again, it excludes events with SFOS lepton pair when its invariant mass is less than 20~GeV or within the $Z$-mass window,
and requires $E_T^{\rm miss} > 50$~GeV.
Only $\tilde \chi_2^0 \tilde \chi_2^0$ intermediate state can produce four charged leptons in our simplified model.
Therefore, this signal region is sensitive only to this class of events.

\section{\texorpdfstring{$p_T$}{pt} and \texorpdfstring{$E_T^{\rm miss}$}{met} distributions and analysis optimisation\label{sec:dist}}

\begin{figure}[t]
\begin{center}
  \subfigure[]{\includegraphics[width=0.49\textwidth]{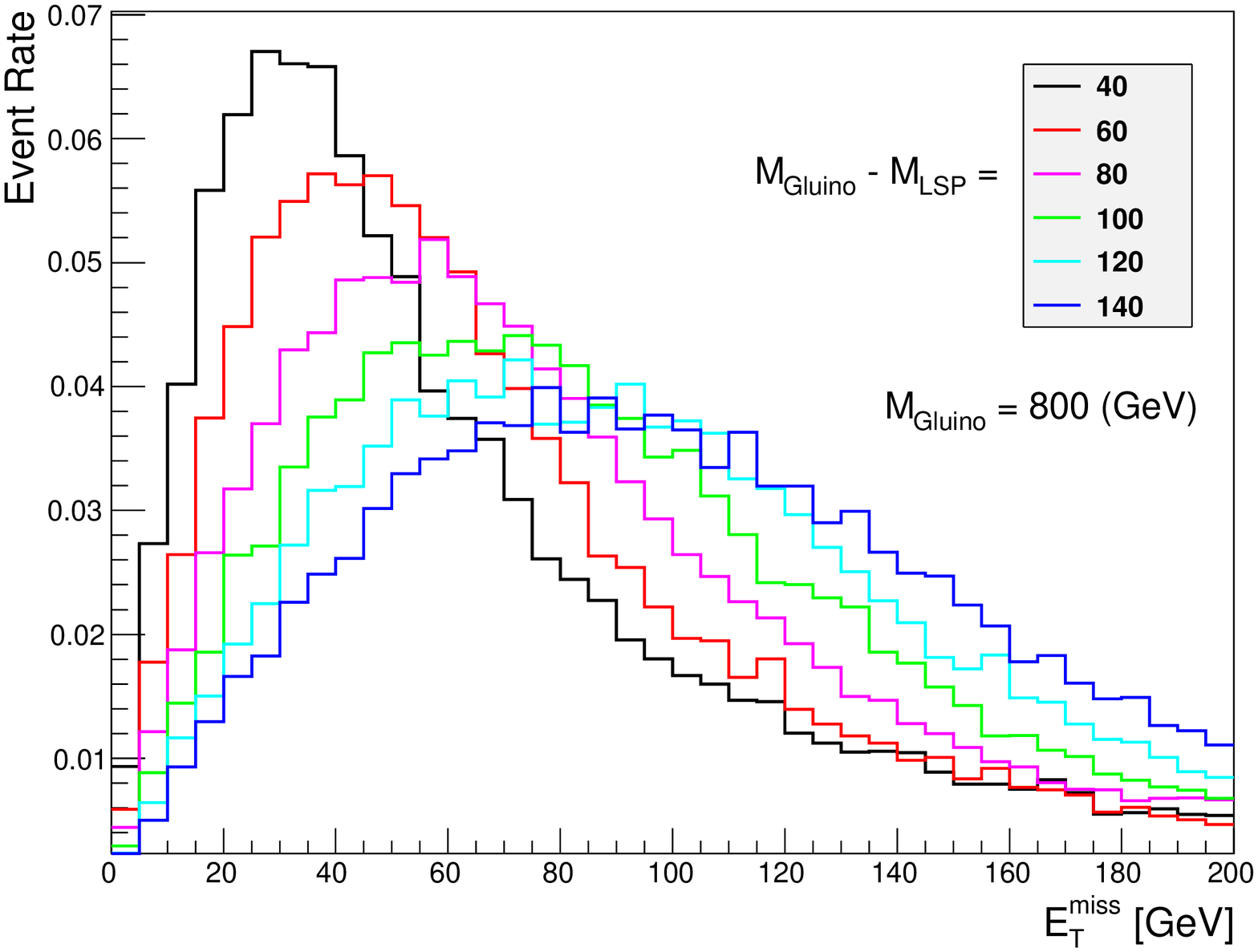}\label{fig:met}}
  \subfigure[]{\includegraphics[width=0.49\textwidth]{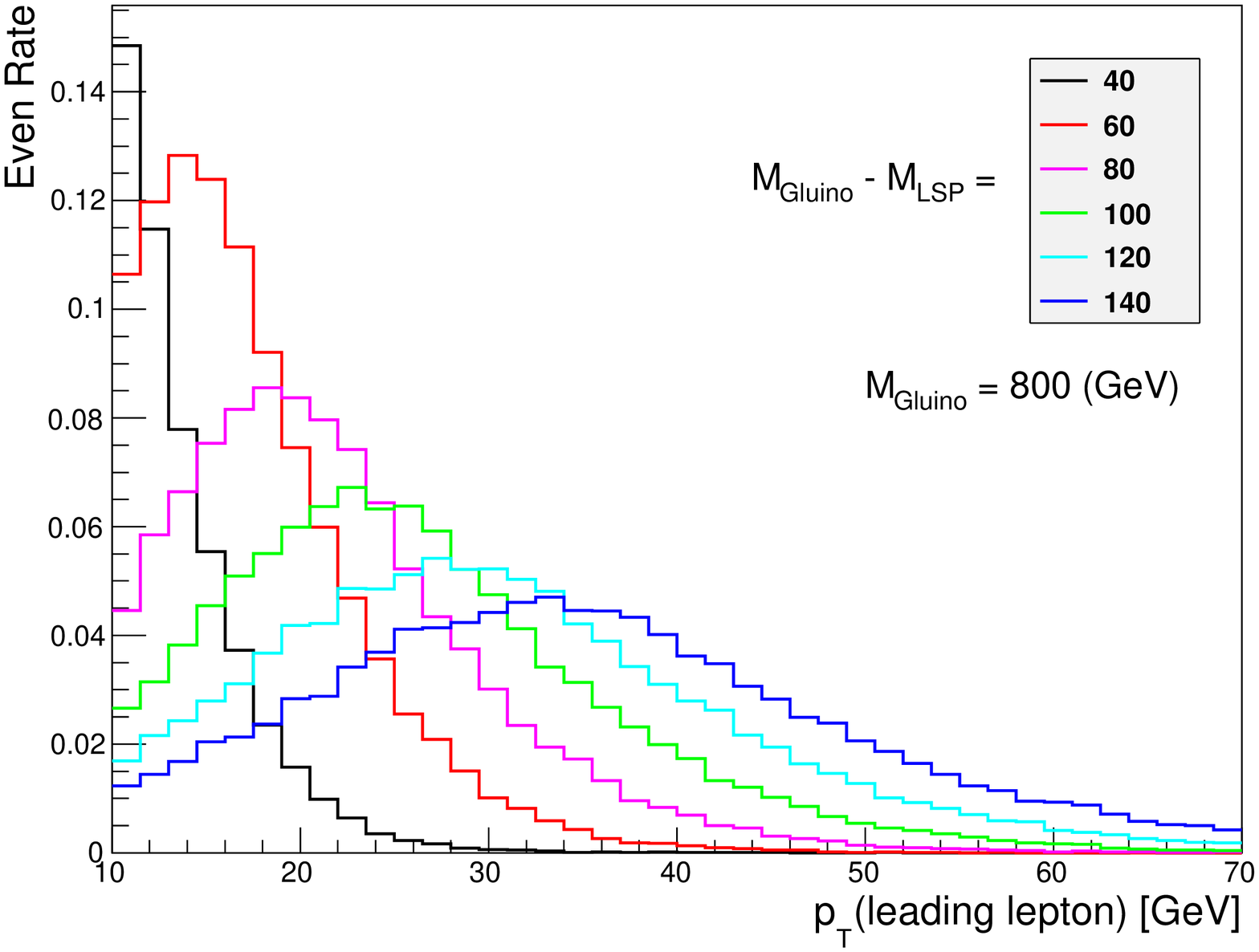}\label{fig:ptlep}}
\caption{(a) Missing energy and (b) transverse momentum of the leading lepton for different values of mass gap, $\Delta m$, between gluino and the LSP.}
\end{center}
\end{figure}

In our simplified model, the source of the missing energy is the two LSPs and zero, one or two neutrinos coming from SUSY cascade decays,
each of which has a typical $p_T$ of 
$\frac{m^2_{\tilde \ell} - m^2_{\tilde B}}{2 m_{\tilde \ell}} \sim \frac{m^2_{\tilde W} - m^2_{\tilde \ell}}{2 m_{\tilde W}} \sim \frac{\Delta m}{4}$.
At the simulation level, $E_T^{\rm miss}$ measurement is based on a vector sum of the transverse momenta of the reconstructed objects.  
Figure~\ref{fig:met} shows the $E_T^{\rm miss}$ distribution for various $\Delta m$ with the fixed gluino mass at 800~GeV.
The distributions peak around $\frac{2}{3} \Delta m$ and exhibit long tails towards the higher $E_T^{\rm miss}$ region. 

The $E_T^{\rm miss}$ distribution of the SM background in the 2SS signal region are provided in figure 1(a) in ref.~\cite{Atlas_2lep}.
The background falls off quickly in the region between 0 to 200~GeV.
From the figure, we can find that the expected background will be reduced by a factor 5, 
if the cut on $E_T^{\rm miss}$ is raised to 140~GeV from 100~GeV in the 2SS signal region.
On the other hand, this modification removes only 70\% of the signal events for $m_{\tilde{g}} = 700$~GeV and $\Delta m = 60$~GeV.  
It is, therefore, expected that if we modify the 2SS signal region by replacing the minimum $E_T^{\rm miss}$ requirement of $100$~GeV with $140$~GeV, 
the sensitivity to the compressed SUSY scenario increases.  
In what follows, we call this modified signal region 2SS+.

Figure~\ref{fig:ptlep} shows a $p_T$ distribution of the leading lepton in the signal.
Again, each histogram corresponds to a different value of $\Delta m$, but the gluino mass is fixed at 800\,GeV.
As can be seen, the distributions have peaks around $\frac{\Delta m}{4}$.  
These distributions can be compared with the leading lepton $p_T$ distribution of the background in the 3LEP signal region 
shown in figure~3 in ref.~\cite{Atlas_3lep_note}. 
The main background in this signal region is coming from diboson ($WZ, ZZ$) production 
and reducible background which originates from an electron from an isolated photon, or an electron or muon coming from decays of heavy flavour mesons. 
The reducible background mainly comes from single top, $t \bar t$ or $WW$ production events.
The leptons from the compressed SUSY signal will typically have low $p_T$, due to the small $\Delta m$, in contrast to the SM background leptons. 
We can read out from the figure that if we employ an additional requirement of $p_T<40$~GeV on the leading lepton,
the background will be reduced by a factor 10.
This can be done by sacrificing just a half of the signal events for $\Delta m = 140$~GeV.  
We use this modified signal region as well as the original signal regions to constrain the compressed SUSY scenario.
Hereafter, we call this signal region 3LEP+.  

\begin{table}[t]
\renewcommand{\arraystretch}{1.5}
\begin{center}
\begin{tabular}{c|c|c|c|c|c|c} \toprule
signal region                                  & 2OS              & 2SS            & 2SS+      & 3LEP                  & 3LEP+     & 4LEP          \\ \hline
leading lepton $p_T$          & $>20(25)$             &  $>20$(25)            &  $>20$(25)    & $>20$(25)            & $>20$(25),$<40$        & $>20$(25)            \\
minimum $E_T^{\mathrm{miss}}$     & $250$       & $100$      & $140$   & $50$                 & $50$     & $50$       \\   
minimum $m_{\ell \ell}$                      &   $12 $       & $12$         & $12$    &$20$   &$20$    &$20$         
\\ \hline
luminosity (${\rm fb^{-1}}$)               & 1.04           & 1.04      & 1.04    &    2.06   &  2.06  & 2.06  \\ 
observed events                & $13$           & $25 $          &  10    & $32$       & $4$       & $0$           \\
expected BG	       & $15.5 \pm 4.0$ & $32.6 \pm 7.9$ &  $6.4 \pm 1.6$    & $26 \pm 5$ & $3.3 \pm 0.6$ & $0.5 \pm 0.8$ \\
$\sigma_{\rm vis}^{\rm bound}$ ($95\%$ CL)            & $9.9 \fb$      & $14.8 \fb$     &  ${10.9} \fb$    & $10.0 \fb$ &  ${ 3.3 \fb}$        & $1.5 \fb$     \\
\bottomrule
\end{tabular}
\end{center}
\renewcommand{\arraystretch}{1.0}
\caption{Signal regions, observed number of events and the corresponding $95\%$ CL upper limit on visible cross section,
see refs.~\cite{Atlas_2lep,Atlas_3lep,Atlas_4lep}. Leading lepton $p_T$ is given separately for muons (electrons).
 \label{tab:signal}}
\end{table}

We summarise the signal regions used in our analysis in table~\ref{tab:signal}. 
The observed events, expected backgrounds and 95$\%$ ${\rm CL_s}$ upper limits on the visible cross section, defined by
cross section times acceptance times efficiency, $\sigma_{\rm vis} = \sigma_{\rm sig} \cdot A \cdot \epsilon$,
are obtained from refs.~\cite{Atlas_2lep,Atlas_3lep,Atlas_4lep}, apart from those for 2SS+ and 3LEP+ signal regions.
For these signal regions, the observed events and expected backgrounds are read off from figure~1(a) 
in ref.~\cite{Atlas_2lep}
and figure~3 in ref.~\cite{Atlas_3lep_note}, respectively, and
the 95$\%$ $\rm CL_s$ limits on the visible cross sections are calculated by following a procedure given in the next section.

\begin{figure}[t]
\begin{center}
  \subfigure[]{\includegraphics[width=0.49\textwidth]{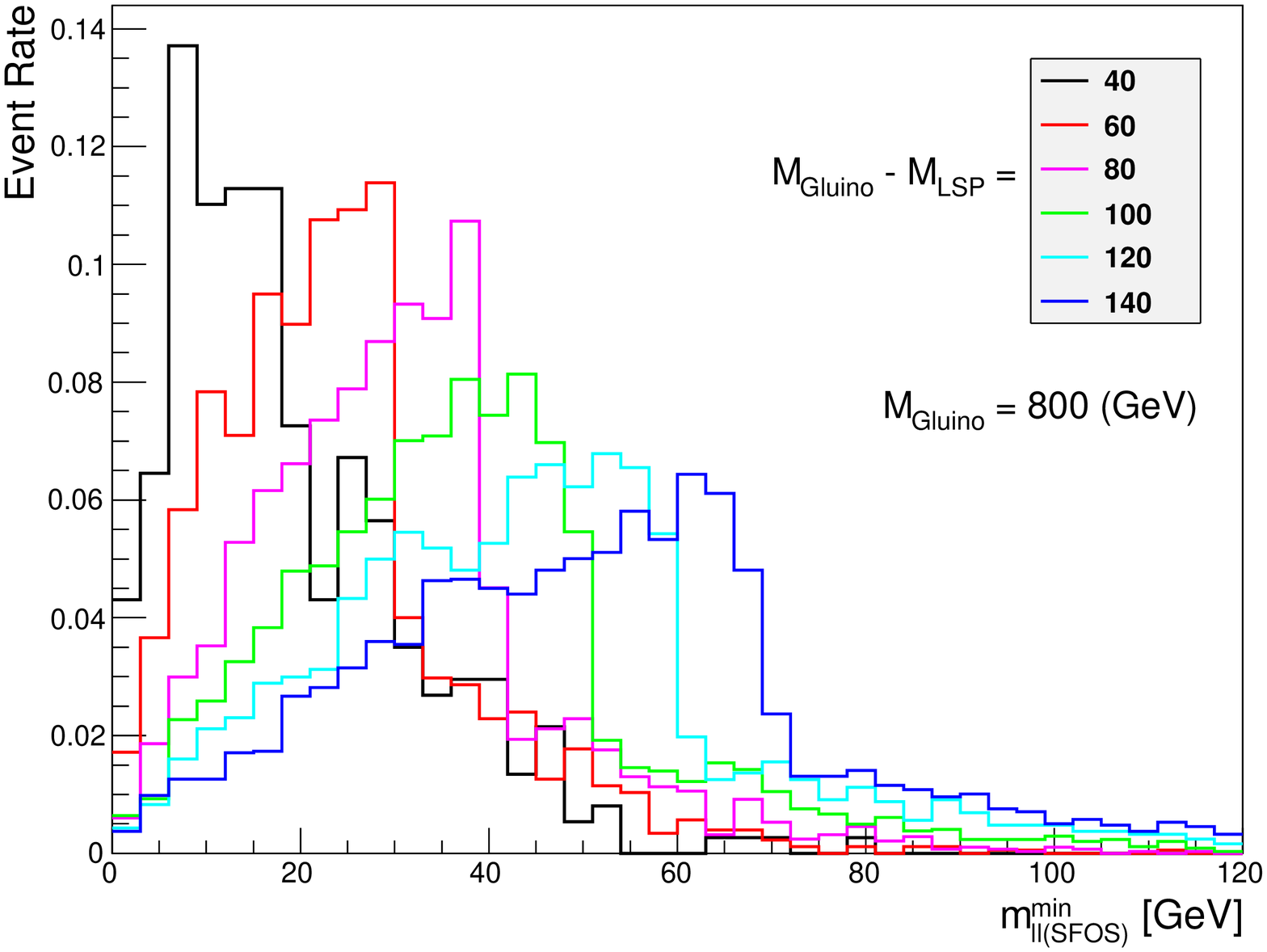}\label{fig:mll}}
  \subfigure[]{\includegraphics[width=0.49\textwidth]{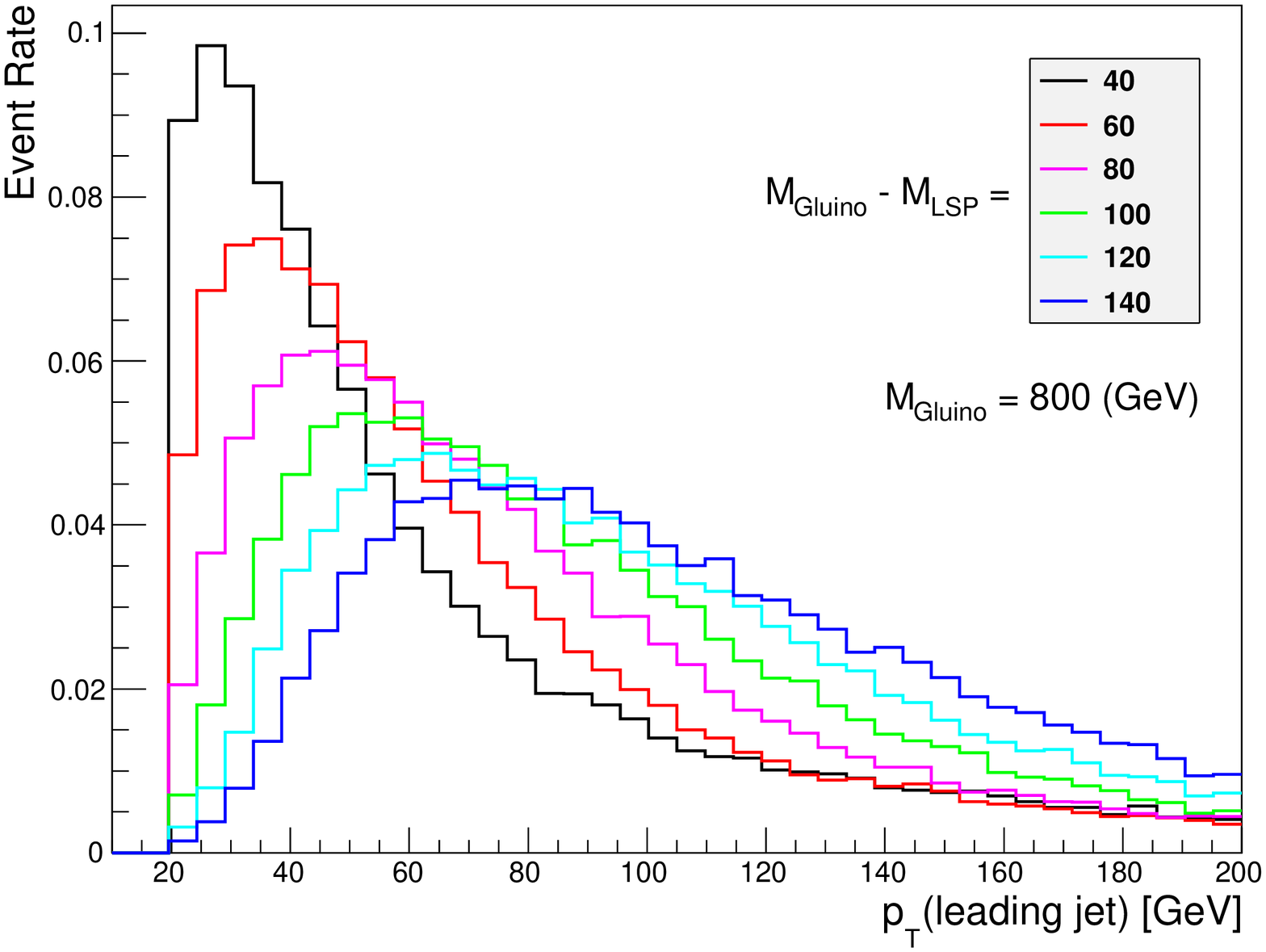}\label{fig:ptjet}}
\caption{(a) Invariant mass of the SFOS lepton pair and (b) transverse momentum of the leading jet for different values of the mass gap, $\Delta m$, between
gluino and the LSP.}
\end{center}
\end{figure}

The dilepton invariant mass cut is as important as the $E_T^{\rm miss}$ and leading lepton $p_T$ cuts  
in constraining the compressed SUSY scenario.
Figure~\ref{fig:mll} shows the invariant mass distributions of a SFOS lepton pairs.        
If more than one SFOS pair is found, only the lowest invariant mass is plotted.
As can be seen, the distributions exhibit the edge structure at 
\begin{equation}
 m_{\ell\ell}^\mathrm{max} = \mneu{2}\sqrt{ \left( 1- \frac{m^2_{\tilde{\ell}}}{\mneu{2}^2} \right)
                                   \left(1-\frac{\mneu{1}^2}{m^2_{\tilde{\ell}}} \right) }\;, \label{eq:invmass}
\end{equation}
which is a maximum for a lepton pair originating from the decay of $\tilde \chi^0_2$. 
The events above $m_{\ell\ell}^\mathrm{max}$ originate from pairing leptons coming from different decay chains.
  
This feature brings a limitation for the analyses using 3LEP, 3LEP+ and 4LEP signal regions, 
which exclude events with SFOS lepton pair with the invariant mass below 20~GeV.
In our simplified model, these signal regions are only sensitive to the events with SFOS lepton pairs coming from $\tilde \chi_2^0$ decays.
Thus, if $\Delta m$ is small and $m_{\ell \ell}^{\rm max}$ becomes less than 20~GeV, 
a contribution to these signal regions will vanish in the limit of a perfect detector resolution.
One can, therefore, expect that these signal regions loose the sensitivity if $\Delta m < 40$\,GeV. 

Note that if $m_{\tilde{\ell}} \simeq \mneu{1} $ or $m_{\tilde{\ell}} \simeq \mneu{2} $ then $m_{\ell\ell}^\mathrm{max} \to 0$ and
$p_T$ of one of the leptons also goes to 0. However, the other lepton would typically be more energetic. In this case one lose sensitivity in 3LEP and 4LEP signal regions, but an enhanced sensitivity could be expected in di-lepton search channels. Finally, if sleptons are heavier than gauginos and only 3-body decays are allowed, both leptons would have similar $p_T$ spectrum leading to a behaviour very similar to our benchmark model. 

Finally we show the $p_T$ distributions of the leading jet in the signal events in Fig.~\ref{fig:ptjet}.
As can be seen, the distributions peak around $\Delta m/2 = m_{\tilde g} - m_{\tilde W}$.
It suggests that information on the mass difference between the strong and weak sectors can be obtained   
if we can measure the shape of the leading jet $p_T$ distribution. 
Feasibility of this measurement will depend on signal-to-background ratio.
A detailed study along this direction
is however beyond the scope of this paper.
Here, we simply stress that the information on the two important mass differences in the compressed SUSY scenario,
$m_{\tilde g} - m_{\tilde W}$ and $m_{\tilde W} - m_{\tilde B}$, is potentially accessible through the distributions of 
leading jet $p_T$ and the invariant mass of SFOS pair.

\section{Obtaining exclusion limits \label{sec:5}}

\subsection{Statistical method \label{subsec:stat}} 

The ATLAS collaboration reported 95\% CL upper limit on the visible cross section in each signal region~\cite{Atlas_2lep,Atlas_3lep,Atlas_4lep}.
One can use those upper limits to assess whether or not a model is excluded at 95\% CL.
In our analysis, we cannot use this simple method, since we want to use new signal regions defined in the previous section.
Instead, we calculate $p$-values by assuming Poisson probability for the number of observed events, 
and construct a $CL_s$ variable~\cite{cls}
including systematic errors on the signal.\footnote{
We use the same method to evaluate a $CL_s$ variable as in ref.~\cite{so10}.  
}

Let $n_{s/b}^{(i)}$ and $\sigma_{s/b}^{(i)}$ be the number of expected events and the systematic error for signal/background in the signal region $i$, respectively.
The number of expected events can then be written as
\begin{equation}
\lambda^{(i)} (\delta_b, \delta_s) = n_b^{(i)} (1 + \delta_b \sigma_b^{(i)}) + n_s^{(i)} (1 + \delta_s \sigma_s^{(i)}) \,, \label{eq:lambda}
\end{equation}
where $\delta_b$ and $\delta_s$ are nuisance parameters, which parametrise the actual size of the systematic errors.
Assuming that the number of observed events follows Poisson distribution 
and systematic errors have Gaussian probability distribution,
the probability of observing $n$ events is given by
\begin{equation}
P(n,n_b^{(i)},n_s^{(i)}) = \int^\infty_{ -1/ \sigma_s^{(i)}} d \delta_s \int^\infty_{ -1/ \sigma_b^{(i)}} d \delta_b 
\frac{e^{-\lambda^{(i)}} (\lambda^{(i)})^n}{n!} e^{-\frac{1}{2}(\delta_s^2 + \delta_b^2)} \,.
\end{equation} 
Note that the lower limits on the integration ranges are set to assure that 
the number of signal and background events are positive.  
If an experiment observes $n_o^{(i)}$ events, 
the $p$-value for the signal plus background hypothesis and that for the background only hypothesis are obtained as
\begin{equation}  
p_{s+b}(n_o^{(i)}) = \sum_{n=0}^{n_o^{(i)}} P(n,n_b^{(i)},n_s^{(i)}) \qquad {\rm and} \qquad
p_{b}(n_o^{(i)})     = \sum_{n=n_o^{(i)}}^{\infty} P(n,n_b^{(i)},0) \,,
\end{equation}  
respectively.
Finally, the $CL_s$ variable is defined as
\begin{equation}  
CL_s = \frac{p_{s+b}}{1 - p_b} \,.
\end{equation}  
The $\rm CL_s$ method sets the exclusion limit at $(1 - CL_s) \cdot 100\,\%$ confidence level. 
Thus, 95\% exclusion is claimed for a model if its $CL_s$ value is less than 0.05.

We have checked that this method can reproduce the reported 95\% CL upper bound on the visible cross section, $\hat \sigma_{\rm vis}^{(i): {\rm bound}}$, 
in each signal region by using the number of expected background events, its error and the number of observed events; see table~\ref{tab:signal}.
In the same way, we compute the visible cross section bounds for the 2SS+ and 3LEP+ signal regions.
We scan $\sigma_{\rm vis}^{(i)}$ substituting $n_s^{(i)}$ for ${\cal L}_{\rm int} \cdot \sigma_{\rm vis}^{(i)}$
and obtain the $95\%$ CL upper bounds by the condition ${\rm CL_s}(\hat \sigma_{\rm vis}^{(i): {\rm bound}}) = 0.05$, where ${\cal L}_{\rm int}$ is an integrated luminosity used in the analysis.
We found $\hat \sigma_{\rm vis}^{(i): {\rm bound}}$ to be 10.9~${\rm fb^{-1}}$ and 3.3~${\rm fb^{-1}}$ for the 2SS+ and 3LEP+ signal regions, respectively.

\subsection{Efficiency}

Before showing the exclusion limits, we discuss the signal efficiency in our simplified model.
Throughout this paper, we simply call signal acceptance times efficiency, $A \cdot \epsilon$, a ``signal efficiency'', which is calculated as
\begin{equation}
A \cdot \epsilon^{(i)} = \frac{{\rm number~of~events~accepted~in~signal~region}~i}{\rm number~of~generated~events} \,.
\label{eq:eff}
\end{equation}   
In the following, the parameter space ($m_{\tilde g}$, $\Delta m$) 
is divided into grids with (100\,GeV, 10\,GeV) step size. 
In each grid point, 50k events are generated in $\Delta m > 60$~GeV region.
When $\Delta m$ is smaller, the numerator of eq.~\eqref{eq:eff} tends to be small, introducing larger MC error.
To avoid a substantial error, 100k events are used in $\Delta m \le 60$\,GeV region.

\begin{figure}[t]
\begin{center}
  \subfigure[]{\includegraphics[width=0.49\textwidth]{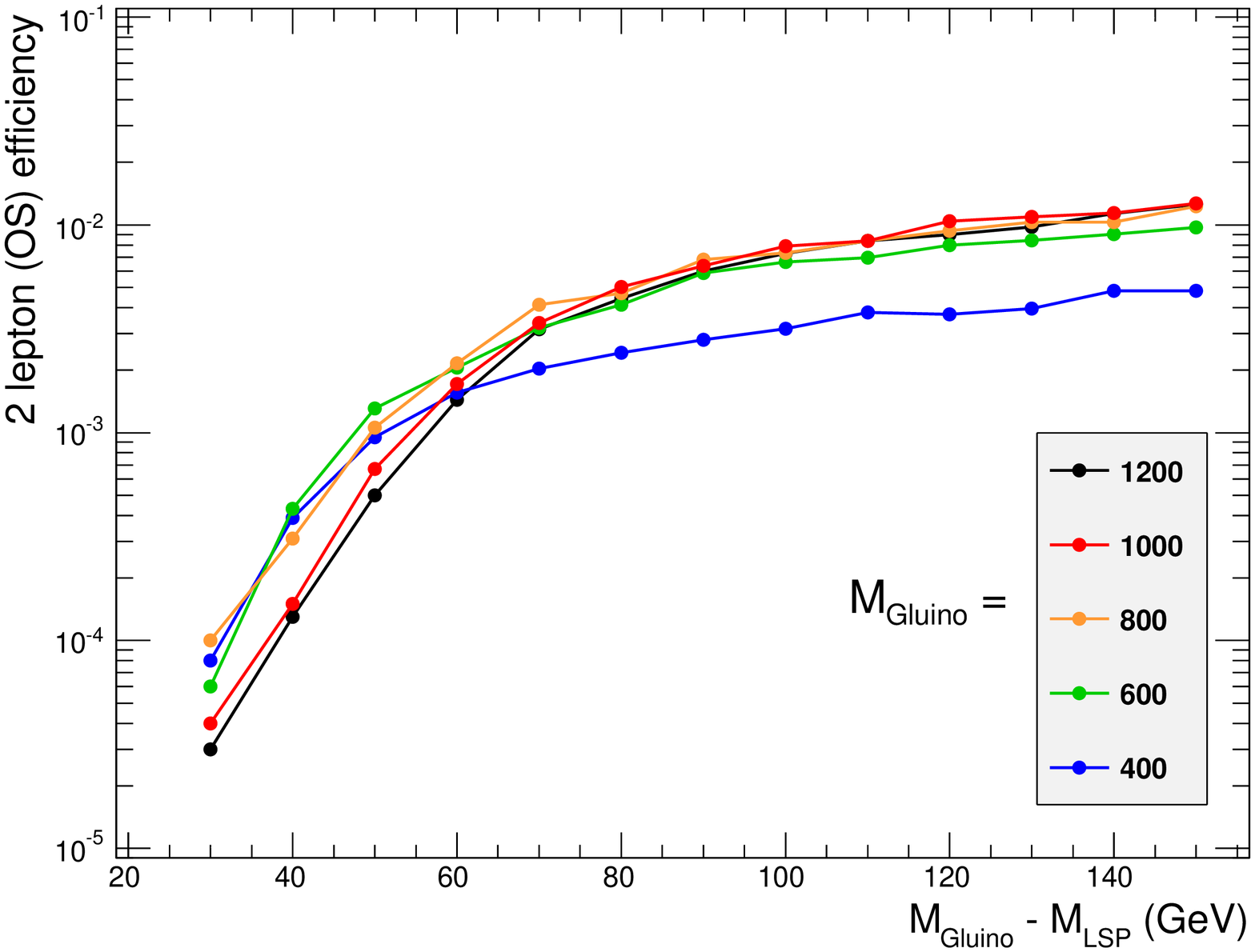}\label{fig:e2os}}
  \subfigure[]{\includegraphics[width=0.49\textwidth]{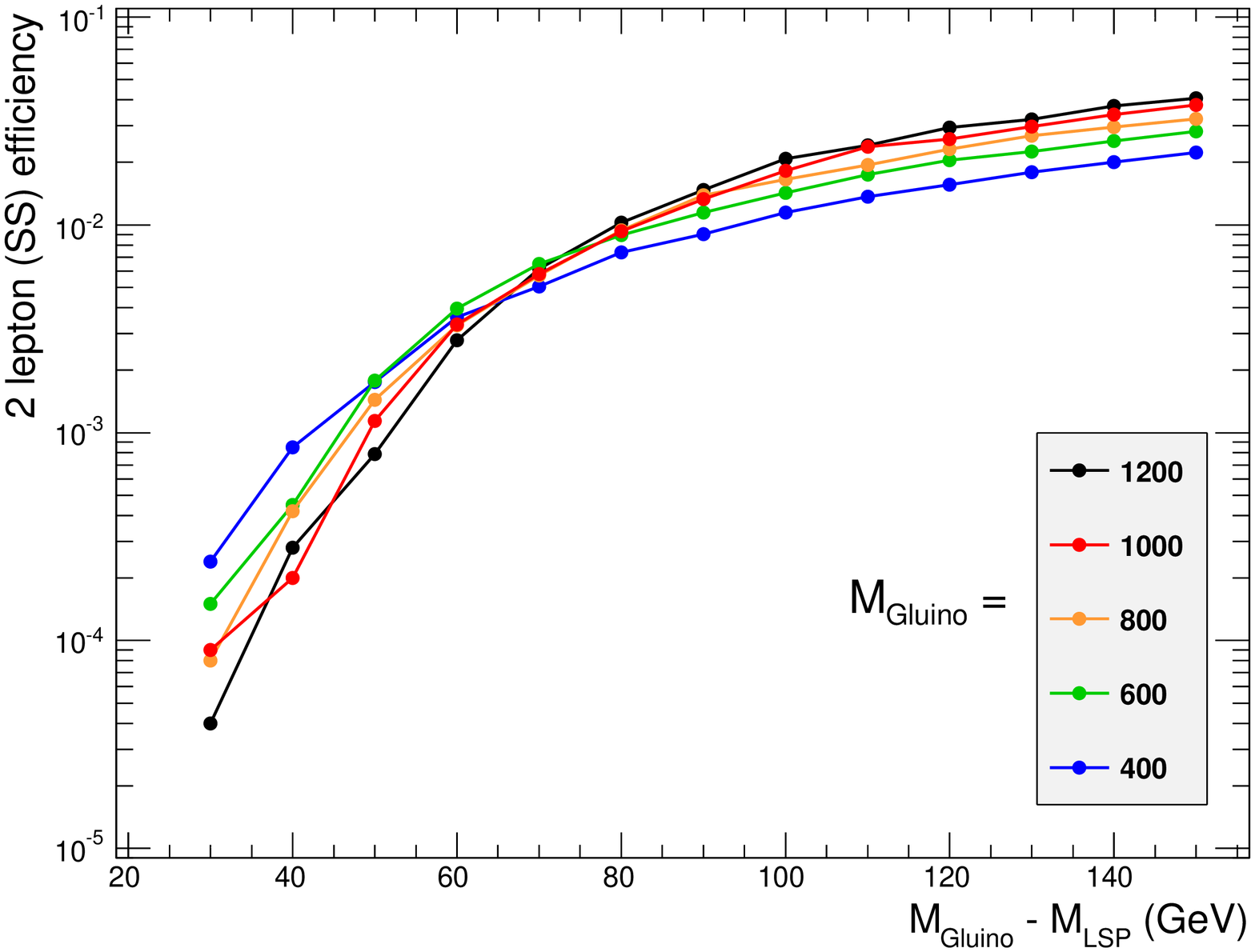}\label{fig:e2ss}}
  \subfigure[]{\includegraphics[width=0.49\textwidth]{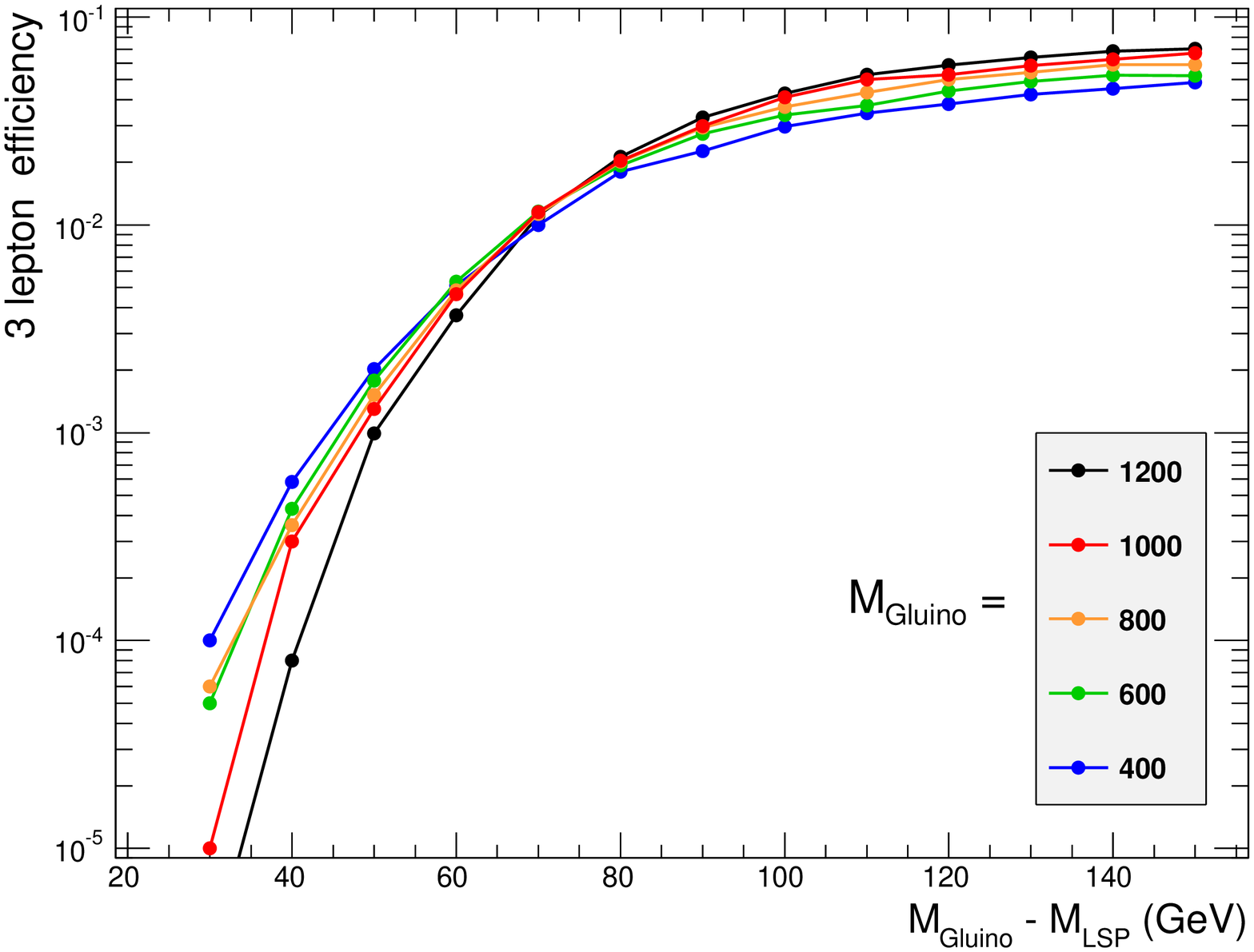}\label{fig:e3lep}}
  \subfigure[]{\includegraphics[width=0.49\textwidth]{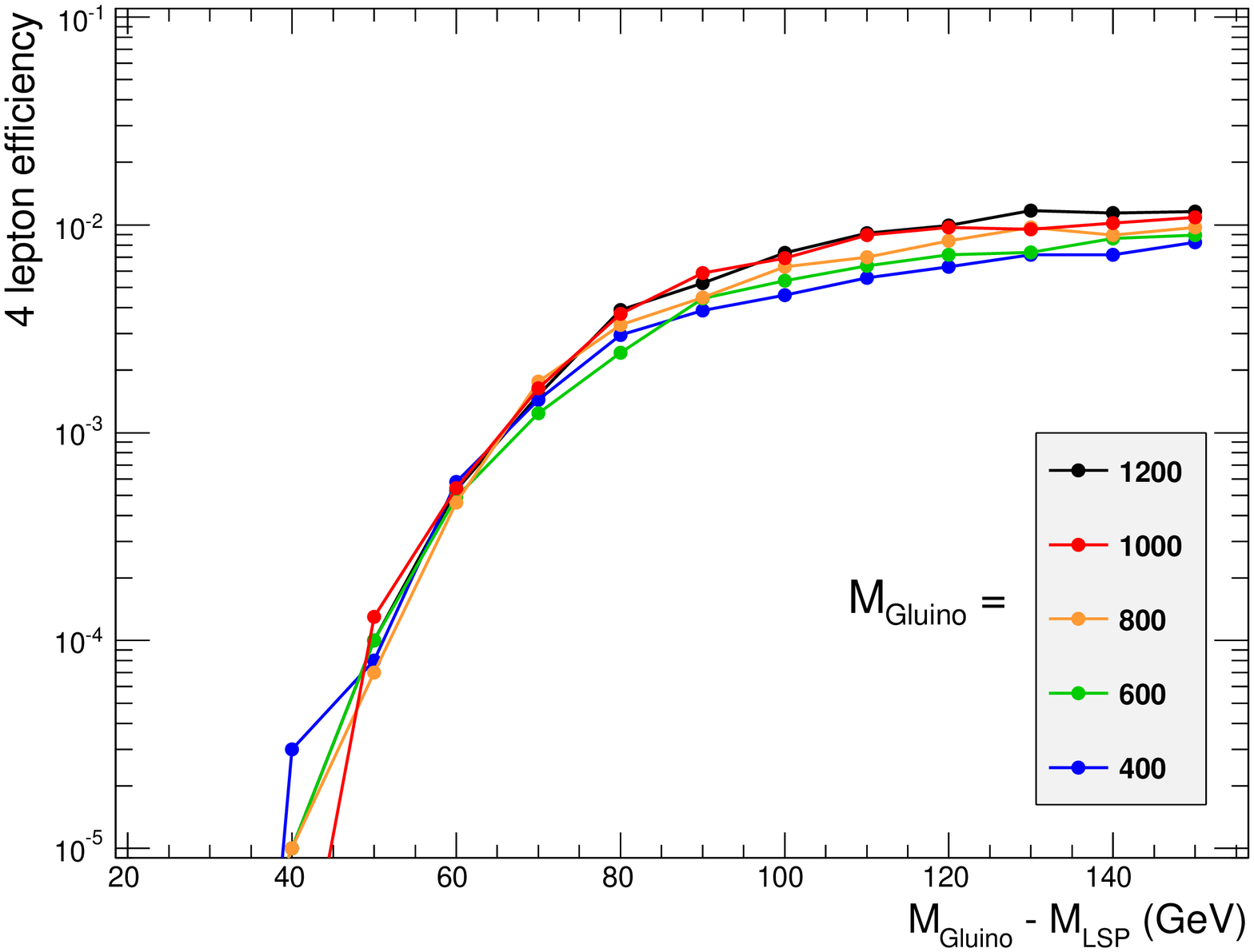}\label{fig:e4lep}}
\caption{Efficiency, eq.~\eqref{eq:eff}, for different signal regions and different gluino masses as a function of the mass splitting, $\Delta m$, between
gluino and the LSP. \label{fig:efficiencies} }
\end{center}
\end{figure}

Figures~\ref{fig:e2os}, \ref{fig:e2ss}, \ref{fig:e3lep} and \ref{fig:e4lep} show the signal efficiency for 2OS, 2SS, 3LEP and 4LEP signal regions, respectively.
The efficiency curves for 2SS+ and 3LEP+ are very similar to those of 2SS and 3LEP signal regions.
The efficiency rapidly drops as $\Delta m$ decreases.
From $\Delta m = 100$~GeV to 40~GeV, the efficiency decrease by 2 or 3 orders of magnitude depending on the signal region.  
Each curve in a plot corresponds to a different choice of the gluino mass.
We can see that the efficiency is less sensitive to the gluino mass than to the mass difference $\Delta m$.
Between $m_{\tilde g} = 1200$~GeV and 400~GeV, the efficiency vary only by a factor of 2 at $\Delta m = 140$~GeV and a factor of 5 at $\Delta m=40$~GeV. This is expected because the scales of lepton $p_T$ and $E_T^{\rm miss}$ are mainly dictated by $\Delta m$.

Nevertheless, the coloured SUSY particle mass, $m_{\tilde g}$ or $m_{\tilde{q}}$, can affect efficiency in the following way.
If $m_{\tilde g}$ is small, $pp$ collisions can produce coloured SUSY particles with higher initial velocities.
In this case, its decay products (jets, leptons, and the LSP) are boosted towards the direction of its velocity 
and the jets and leptons are more likely to be collimated.
Those hadronic activity around the lepton reduces the efficiency of lepton isolation and 
has some negative effect on the signal efficiency.
From figure~\ref{fig:efficiencies}, we can see that the signal efficiency becomes smaller as $m_{\tilde g}$ decreases in the large $\Delta m$ region. 
On the other hand, in the small $\Delta m$ region, this feature gives the opposite effect.
In this region, the signal efficiency drops because the lepton $p_T$ and $E_T^{\rm miss}$ tend to be too small
to pass the high $p_T$ and $E_T^{\rm miss}$ cuts.
The boost effect, in this case, helps leptons and LSPs to have large enough momenta
to pass those cuts.

\subsection{95\% exclusion limits} 
 
\begin{figure}[t]
\begin{center}
  \subfigure[]{\includegraphics[width=0.49\textwidth]{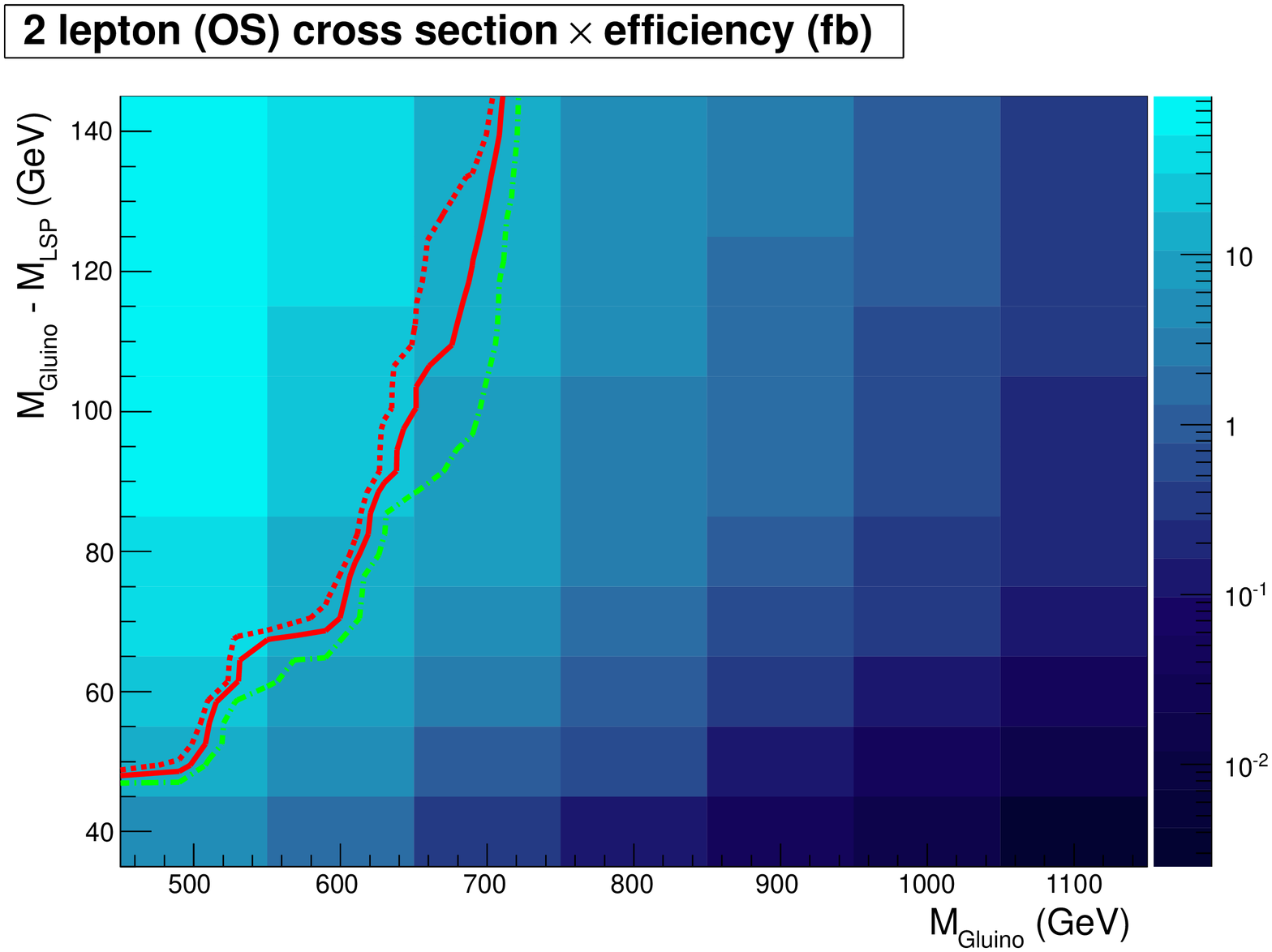}\label{fig:eff_2os}}
  \subfigure[]{\includegraphics[width=0.49\textwidth]{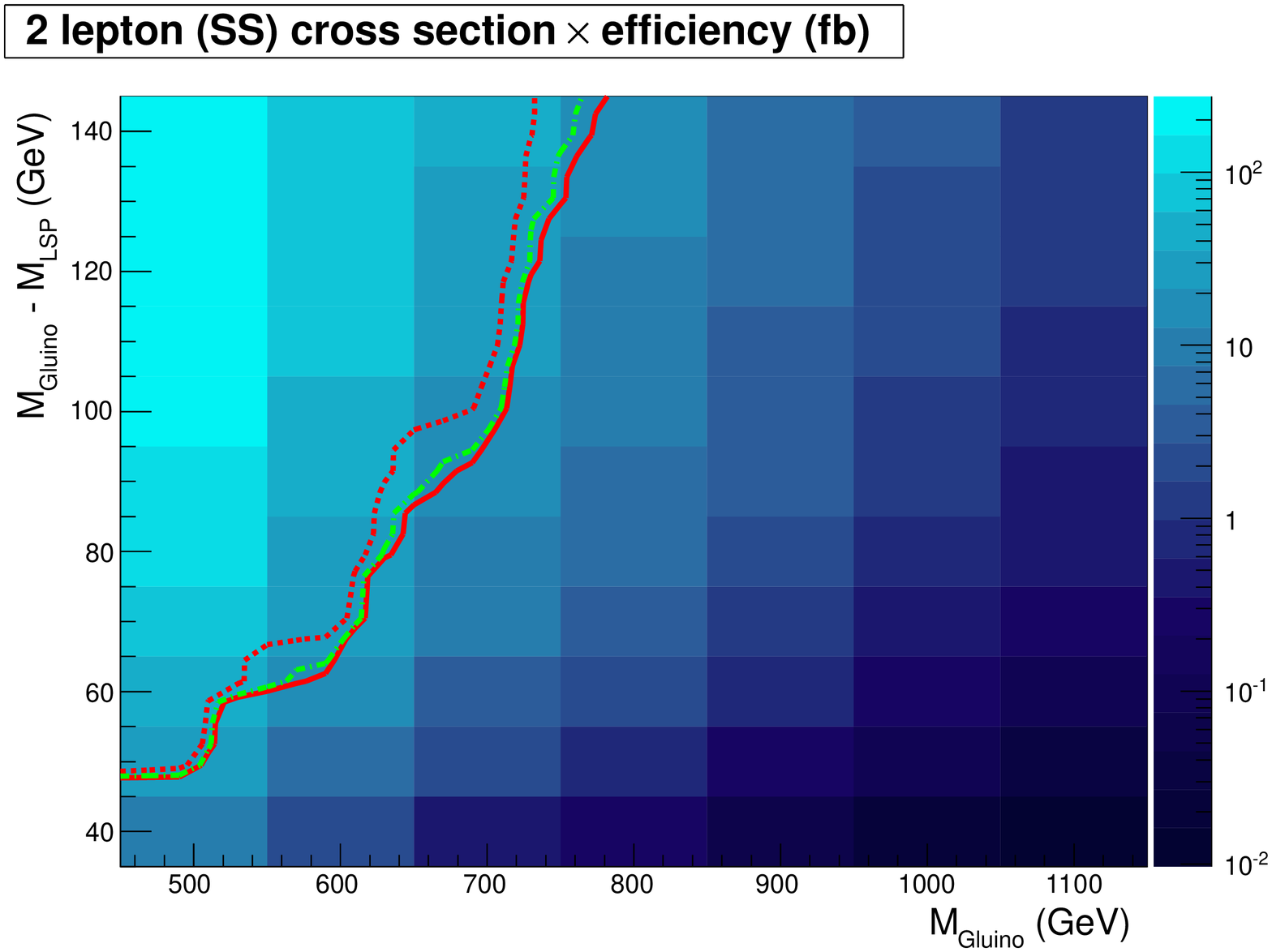}\label{fig:eff_2ss}}
  \subfigure[]{\includegraphics[width=0.49\textwidth]{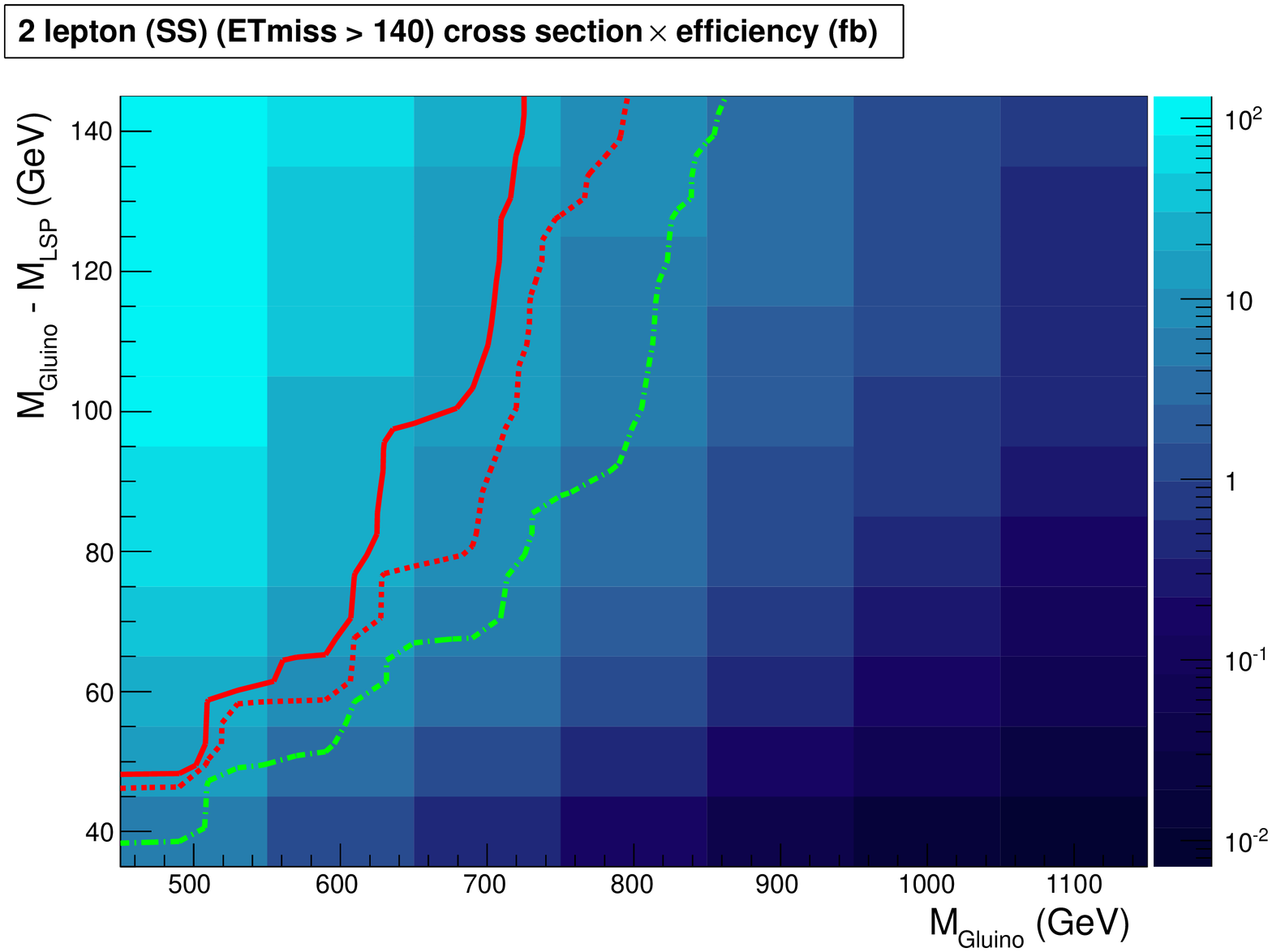}\label{fig:eff_2ss+}}
  \subfigure[]{\includegraphics[width=0.49\textwidth]{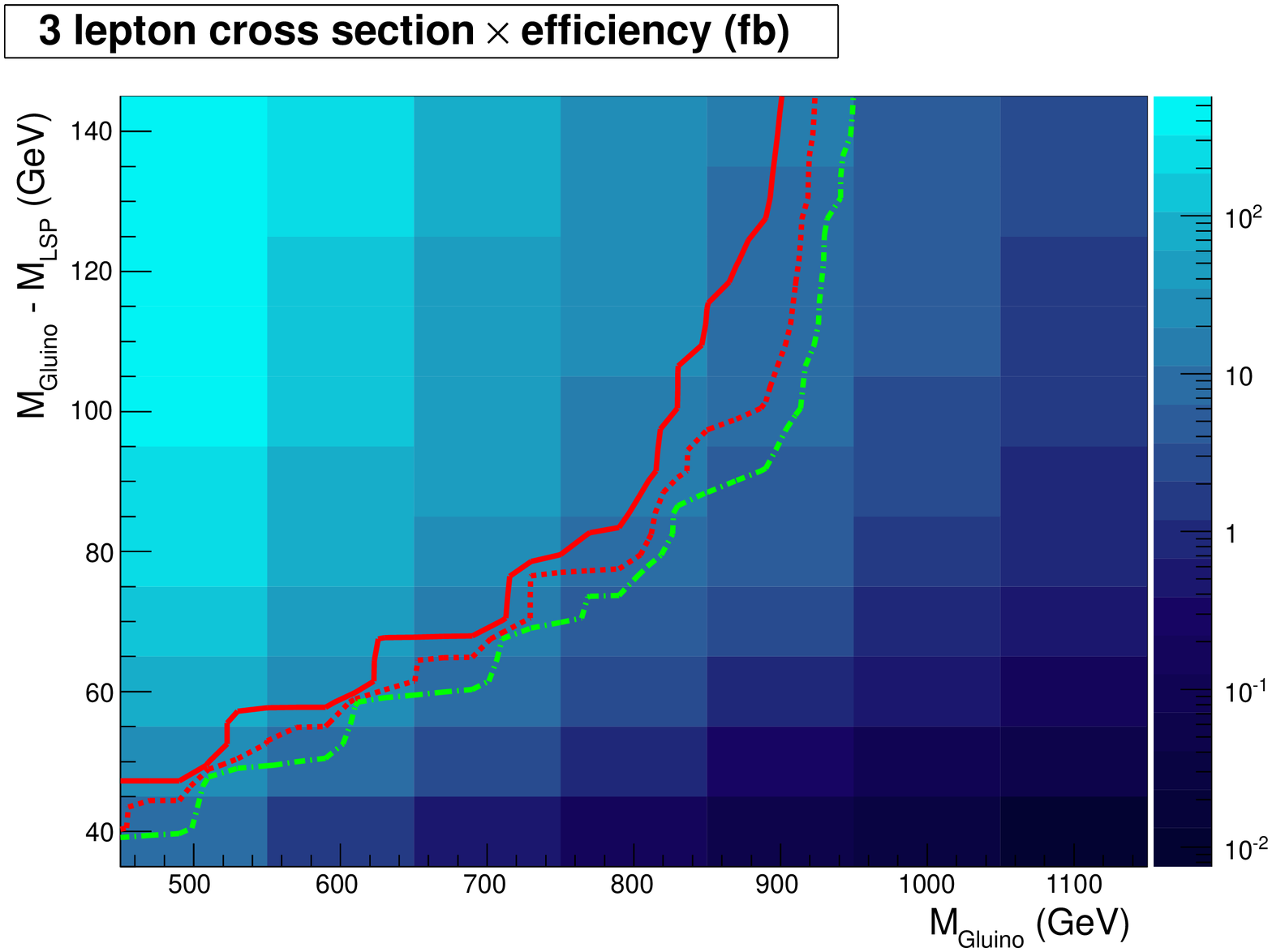}\label{fig:eff_3lep}}
  \subfigure[]{\includegraphics[width=0.49\textwidth]{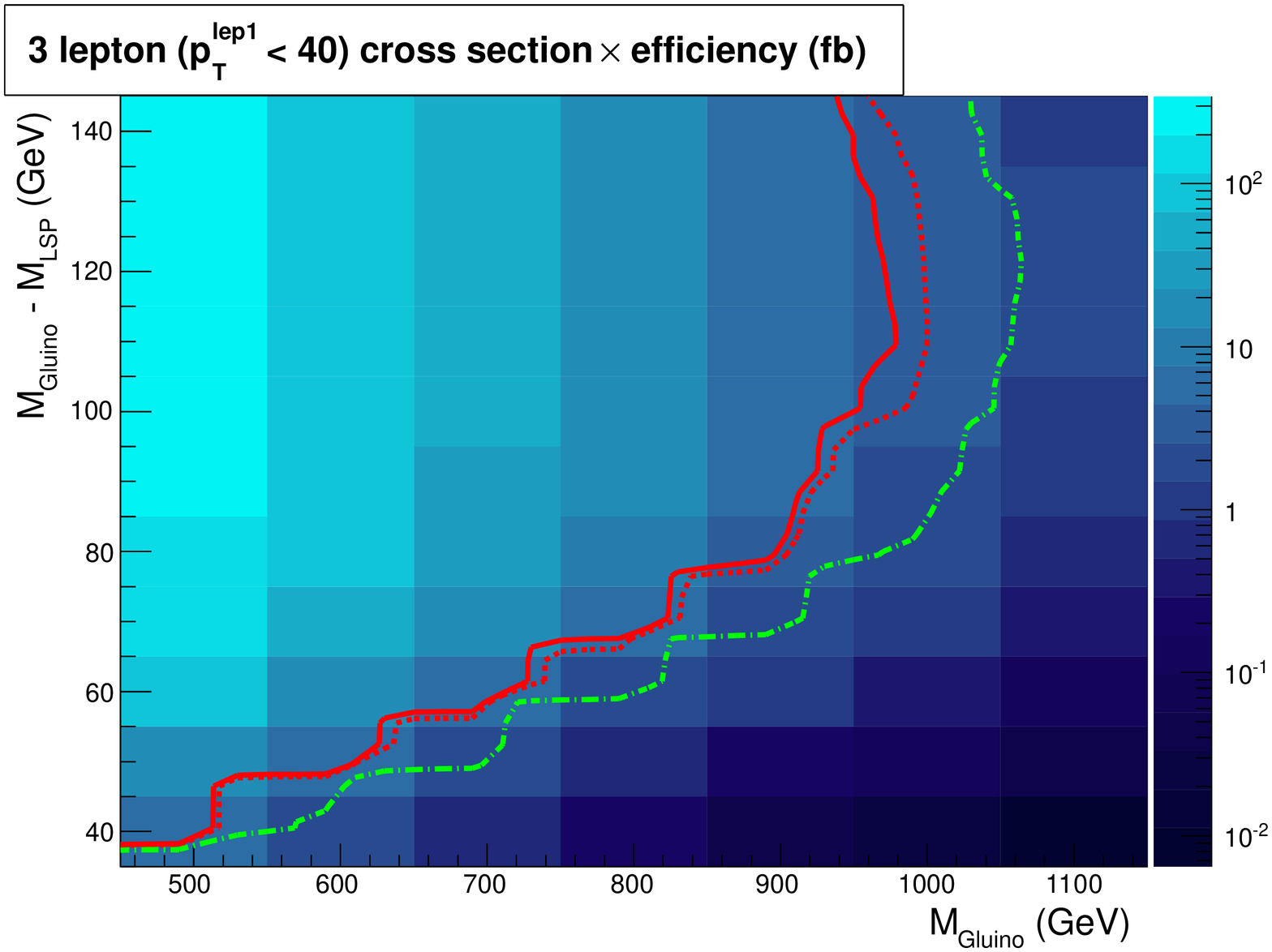}\label{fig:eff_3lep+}}
  \subfigure[]{\includegraphics[width=0.49\textwidth]{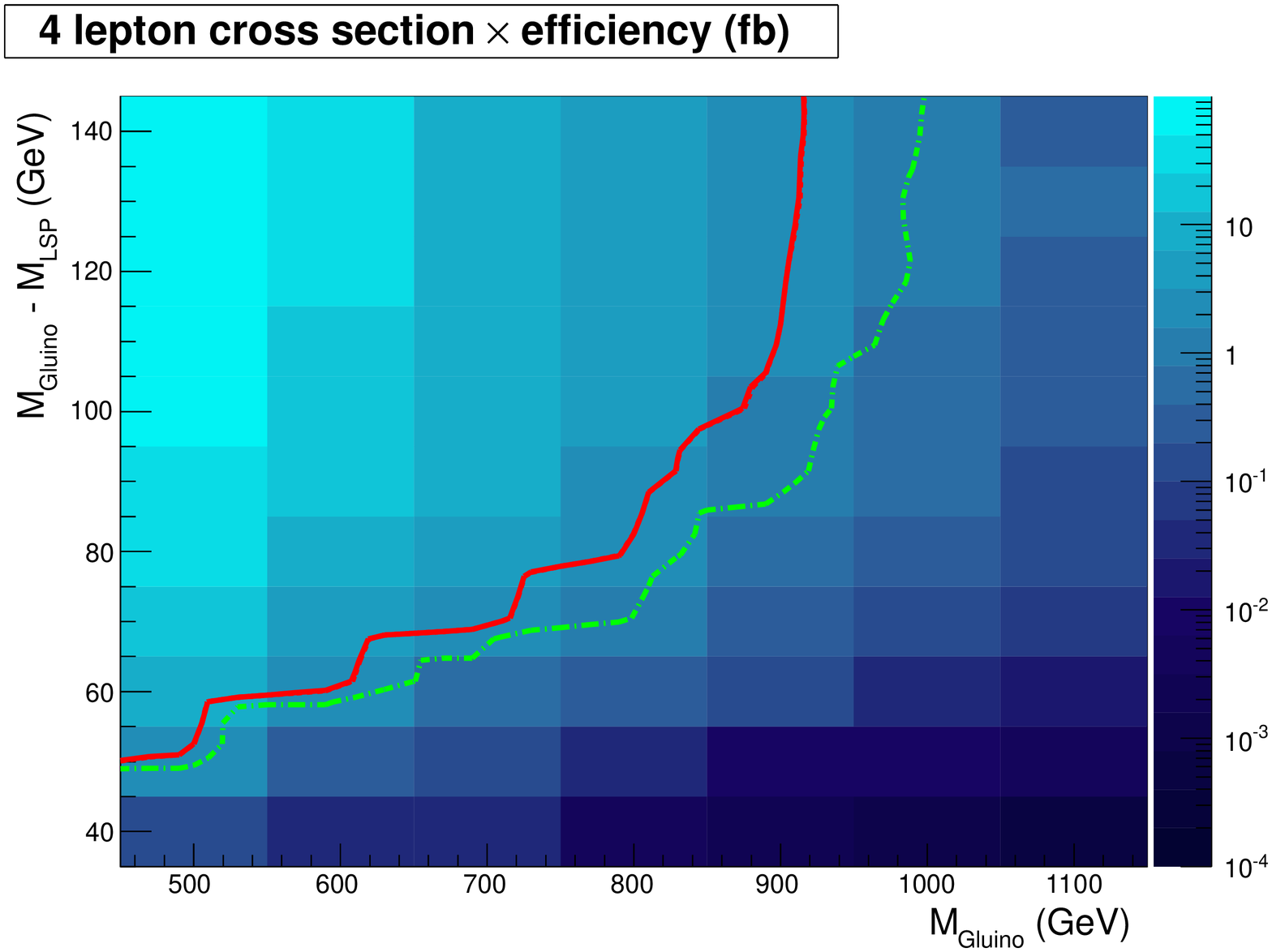}\label{fig:eff_4lep}}
\caption{Exclusion plots for the studied signal regions. The colours denote excluded cross section and the lines $95\%$ CL exclusion limits:
observed (red solid), expected (red dashed, at $\mathcal{L}_{\rm int} = 1 \fb^{-1}$ or $\mathcal{L}_{\rm int} = 2.06 \fb^{-1}$, see table~\ref{tab:signal}), 
and expected with $\mathcal{L}_{\rm int} = 5.25 \fb^{-1}$ (green dashed).\label{fig:exclusion}}
\end{center}
\end{figure}

Six plots in figure~\ref{fig:exclusion} show the visible cross section in each signal region (\ref{fig:eff_2os} 2OS, \ref{fig:eff_2ss} 2SS, \ref{fig:eff_2ss+} 2SS+, \ref{fig:eff_3lep} 3LEP, \ref{fig:eff_3lep+} 3LEP+, \ref{fig:eff_4lep} 4LEP) 
in ($m_{\tilde g}$, $\Delta m$) plane.
In the same plots, we superimpose the 95$\%$ observed (red solid) and expected (red dashed) exclusion limits with the corresponding luminosities 
used in the analyses; see table~\ref{tab:signal}.
The green dashed curves represent the expected 95$\%$ exclusion limits with the integrated luminosity of 5.25\,${\rm fb^{-1}}$. 
The visible cross section is calculated by $\sigma_{\rm vis}^{(i)} = \sigma_{\rm tot} \cdot A\cdot \epsilon^{(i)}$, where $\sigma_{\rm tot}$ is
the total SUSY production cross section.
For $\sigma_{\rm tot}$, we use the next-to-leading order cross section calculated using \texttt{Prospino~2.1}~\cite{Beenakker:1996ch}. 

As can be seen, the visible cross section strongly depends on both $m_{\tilde g}$ and $\Delta m$.
This dependence is approximately factorisable as 
$\sigma_{\rm vis}(m_{\tilde g},\Delta m) = \sigma_{\rm tot}(m_{\tilde g}) \cdot A\cdot \epsilon(\Delta m)$.
$\sigma_{\rm tot}$ is almost exclusively determined by $m_{\tilde g}$, because the coloured SUSY particle production
dominate the total SUSY production in the compressed SUSY scenario.
On the other hand, $A\cdot \epsilon$ is dependent mostly on $\Delta m$, as we have seen in the previous subsection.   

The $95\%$ exclusion limits are calculated using the ${\rm CL_s}$ method described in subsection~\ref{subsec:stat}.
We simply use the signal systematic error of 20$\%$ ($\sigma_{s} = 0.2$), cf.\ eq.~\ref{eq:lambda}, for each of the signal regions  
across the parameter space.
In calculating the expected limit for an integrated luminosity $\mathcal{L}_{\rm int} = 5.25\ {\rm fb^{-1}}$,
we conservatively use the same background systematic error as the one used in the lower luminosity analyses by ATLAS.
If the background systematic error is improved in the updated analysis with $\mathcal{L}_{\rm int} = 5.25\ {\rm fb^{-1}}$, 
the expected exclusion limits will be slightly improved.

\begin{figure}[t]
\begin{center}
 \subfigure[]{\includegraphics[width=0.49\textwidth]{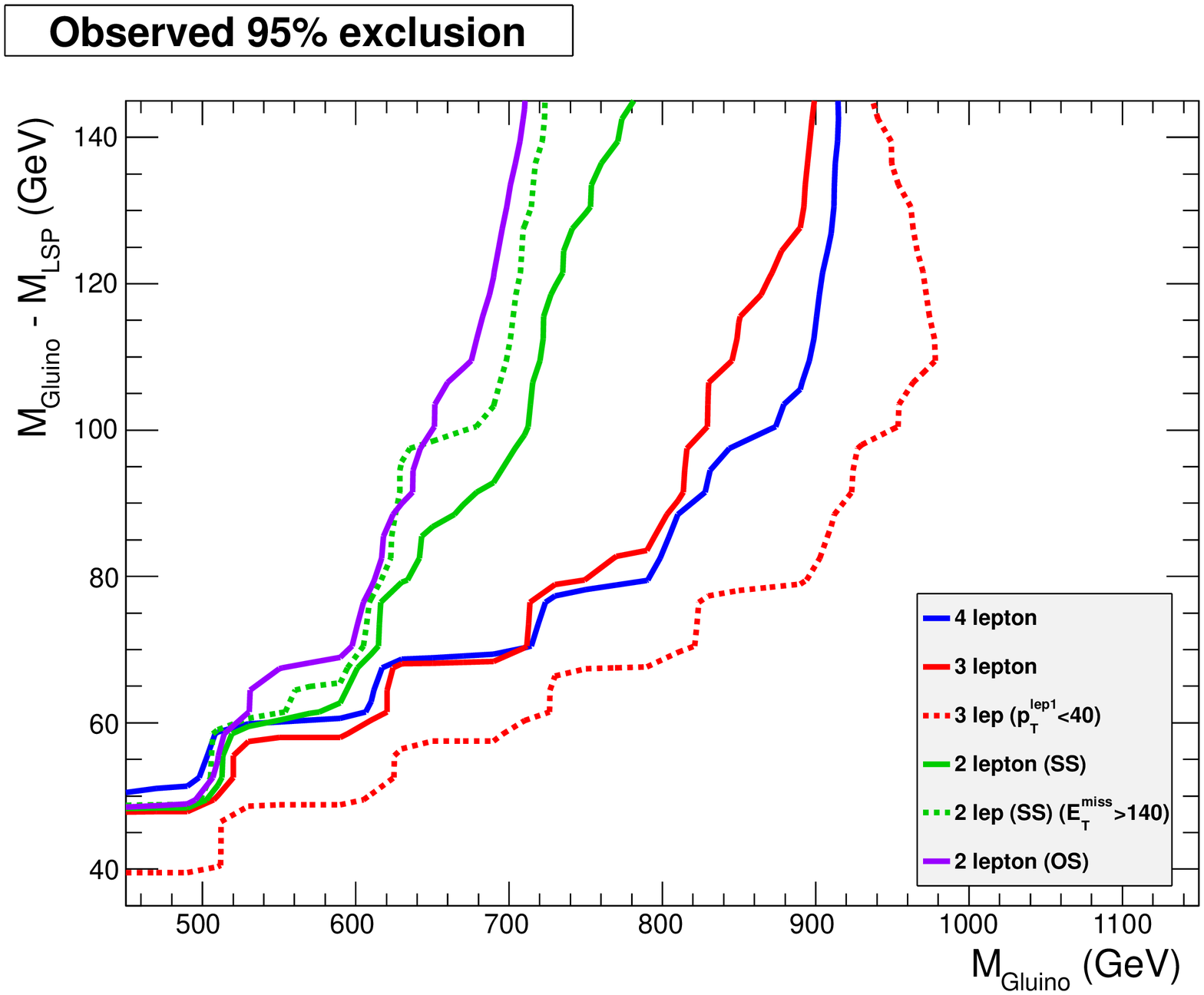}\label{fig:ex_ob}}
 \subfigure[]{\includegraphics[width=0.49\textwidth]{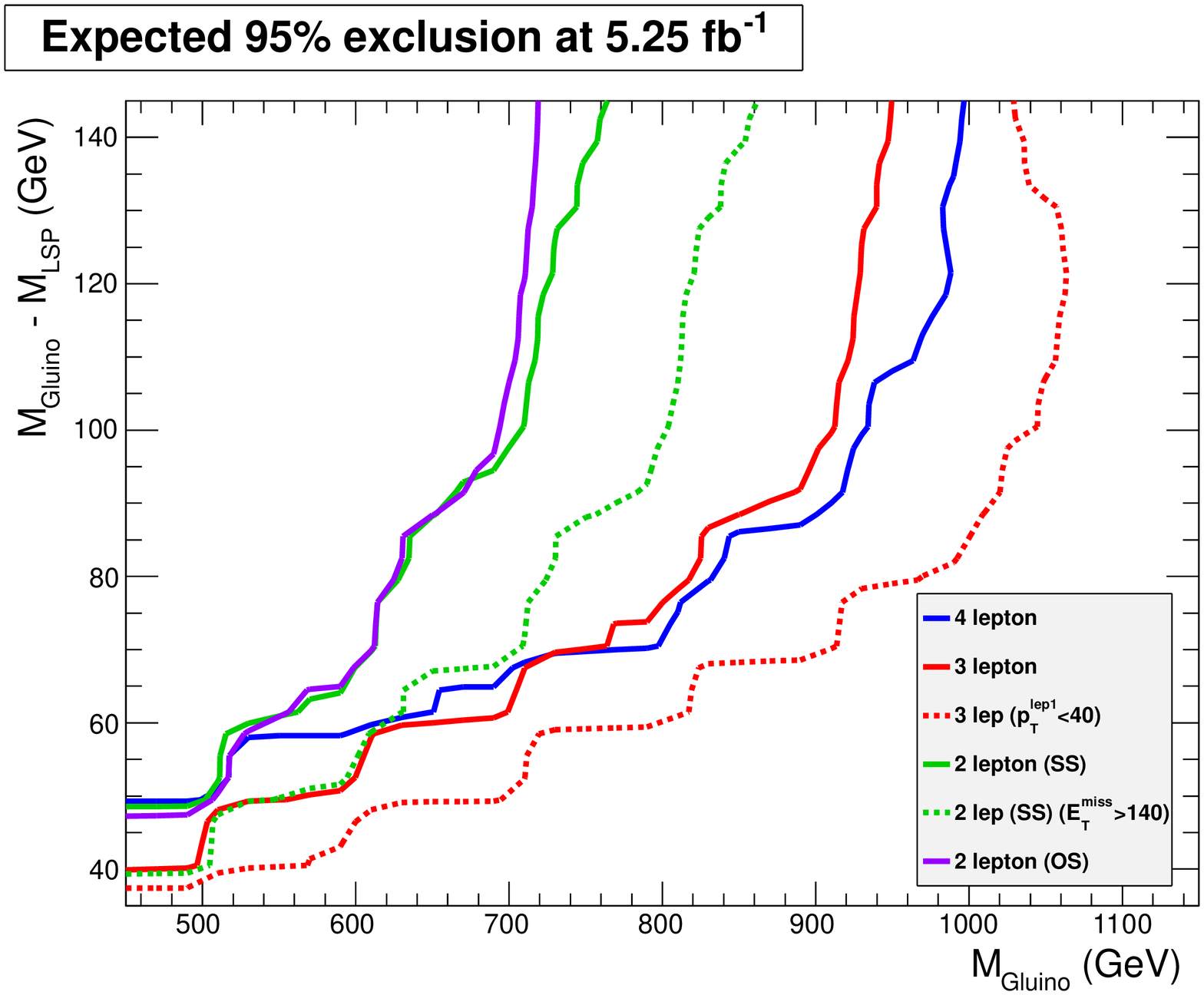}\label{fig:ex_ex}}
\caption{Exclusion limits (a) observed ($\mathcal{L}_{\rm int} =$ 1--2$\fb^{-1}$) and (b) expected for $\mathcal{L}_{\rm int} = 5.25 \fb^{-1}$.\label{fig:exclsum} }
\end{center}
\end{figure}

We summarise the observed and expected (with $\mathcal{L}_{\rm int} = 5.25\ {\rm fb^{-1}}$) exclusion limits obtained from various signal regions
in figures~\ref{fig:ex_ob} and \ref{fig:ex_ex}, respectively.
As can be seen, 4LEP signal region provides the most stringent limit among the original analyses,
however the modified 3LEP+ signal region sets even better exclusion limit than the 4LEP signal region 
in the ($m_{\tilde g}$, $\Delta m$) parameter plane.
The exclusion limits turn out to be quite strong:
the gluino and squark masses below 900~GeV are excluded for $\Delta m > 80$~GeV
and the 600~GeV gluino/squark mass is excluded for $\Delta m > 50$~GeV.
The analyses do not exclude any gluino/squark masses for $\Delta m$ below $40$~GeV.
This is expected since the kinematical upper bound on the $\tilde \chi_2^0$-originated SFOS lepton pair 
and a typical size of lepton $p_T$ become smaller than their cut threshold values,
and the analyses lose sensitivity in this region as discussed in section~\ref{sec:dist}.    
For the di-lepton analysis, our modified signal region 2SS+ does not improve the observed exclusion limit with respect to the original 2SS signal region.
This is because of a $2 \sigma$ excess in the 2SS+ signal region observed by ATLAS; see table~\ref{tab:signal}.
From figure~\ref{fig:ex_ex} we can see that the 2SS+ signal region indeed has a better sensitivity against the 2SS signal region,
where the limit on the coloured SUSY particle mass is extended by about 100\,GeV in $\Delta m > 70$~GeV region. Finally, we have checked that replacing the sequence of two-body gaugino decays by three-body decays (ie.\ by making sleptons heavier than gauginos) yields practically the same exclusion limits.

\section{Signal decomposition and visible cross section reconstruction \label{sec:6}}

The exclusion limits obtained in the previous section rely on the details of the simplified model.
For the models with different mass spectrum and branching ratios, the limit is not applicable. 
Recently, an application of a simplified model limits to constrain other models has been discussed~\cite{simp}.
If a model contains a specific event topology for which visible cross section upper bound is known from a simplified model study, we can constrain the model by imposing that limit on the partial cross section of that topology.   

The event topologies of our simplified model can be decomposed into five classes each of which has different intermediate weakino states:
\begin{center}
\begin{description}
\item[(i)]  $\tilde \chi_1^{\pm} \tilde \chi_1^{\pm}$~$\Longrightarrow$~2OS,~2SS; 
\item[(ii)]  $\tilde \chi_2^0 \tilde \chi_1^0$~$\Longrightarrow$~2OS; 
\item[(iii)]  $\tilde \chi_1^{\pm} \tilde \chi_2^0$~$\Longrightarrow$~3LEP~(2OS,~2SS);
\item[(iv)]  $\tilde \chi_2^0 \tilde \chi_2^0$~$\Longrightarrow$~4LEP~(3LEP, 2OS, 2SS); 
\item[(v)]  $\tilde \chi_1^0 \tilde \chi_1^0$, $\tilde \chi_1^{\pm} \tilde \chi_1^0$~$\Longrightarrow$~less than 2 leptons.
\end{description}
\end{center}
The last class does not produce two or more leptons in the final state and is therefore irrelevant for the multi-lepton searches.
The $\tilde \chi_1^{\pm} \tilde \chi_1^{\pm}$ and $\tilde \chi_2^0 \tilde \chi_1^0$ classes are expected to have two isolated leptons
and the former can have either OS and SS lepton pair while the latter only has OS lepton pair. 
The $\tilde \chi_1^{\pm} \tilde \chi_2^0$ state can produce up to three leptons but can also contribute to 2OS and 2SS signal regions
if one of the leptons fails to satisfy reconstruction requirements.
The $\tilde \chi_2^0 \tilde \chi_2^0$ class can contribute to all the signal regions once the lepton acceptance and isolation efficiency are taken into account.   

The visible cross section in our simplified model can be decomposed as
\begin{equation}
\sigma_{\rm vis}^{(i)} = \sum_{a,b,\gamma,\delta} \sigma_{a b} \cdot B_{a \to \gamma} \cdot B_{b \to \delta} \cdot 
B_{\gamma} \cdot B_{\delta} \cdot
A \cdot \epsilon_{\gamma \delta}^{(i)}\;,
\label{eq:dec}
\end{equation}
where $a$ and $b$ denote the particles produced in $pp$ collision ($ a,b = \tilde g, \tilde q$) and
$\gamma$ and $\delta$ denote intermediate weakinos ($\gamma,\delta = \tilde \chi_1^{\pm}, \tilde \chi_2^{0}, \tilde \chi_1^{0}$).
$\sigma_{ab}$ is the cross section of $pp \to ab$ production process 
and $B_{a \to \gamma}$ and $B_{\gamma}$ are branching ratios of the corresponding decays.
Note that $B_{\gamma}$ is $Br(\tilde \chi_1^{\pm} \to \ell^{\pm} \nu_\ell \tilde \chi_1^0)$ for $\gamma =\tilde \chi_1^{\pm}$ and
$Br(\tilde \chi_2^{0} \to \ell^{\pm} \ell^{\mp} \tilde \chi_1^0)$ for $\gamma =\tilde \chi_2^0$.
The $A\cdot \epsilon_{\gamma \delta}^{(i)}$ represents the efficiency in the signal region $i$ for $\gamma \delta$ event class. The efficiencies for different topologies and signal regions in our simplified model are collected in table~\ref{tab:eff}.

If a model has cascade decay chains as in eq.~\eqref{eq:br}, a contribution to the visible cross section from classes (i)--(iv) can be calculated as:
\begin{equation}   
\hat \sigma_{\rm vis}^{(i)} 
=\sum_{a,b,\gamma,\delta}\hat \sigma_{a b} \cdot \hat B_{a \to \gamma} \cdot \hat B_{b \to \delta} \cdot 
\hat B_{\gamma} \cdot \hat B_{\delta} \cdot
A \cdot \epsilon_{\gamma \delta}^{(i)} \,,
\label{eq:svis}
\end{equation}  
where $\hat \sigma_{ab}$ and $\hat B_{(\ldots)}$ represent cross sections and branching ratios calculated in the given model. If the model features similar kinematics to our simplified model, the relevant efficiency can be found in table~\ref{tab:eff} and no dedicated MC simulation is required.
Therefore the model is excluded if
\begin{equation}
  \hat \sigma_{\rm vis}^{(i)} > \sigma_{\rm vis}^{(i):{\rm bound}} \,,
\label{eq:svis2}
\end{equation}  
where $\sigma_{\rm vis}^{(i):{\rm bound}}$ is the reported model-independent upper bound shown in table~\ref{tab:signal}.

In the compressed SUSY scenario, $A\cdot\epsilon_{\gamma \delta}^{(i)}$ depends mainly on $\Delta m$ in the first approximation.
Therefore, in table~\ref{tab:eff} we list the decomposed efficiencies $A\cdot \epsilon_{\gamma \delta}^{(i)}$ for each $\Delta m$ for $m_{\tilde g} = 800$~GeV.
For different values of $m_{\tilde g}$, the efficiencies vary by about factor of $5$ in $\Delta m < 40$~GeV and factor of 2 in $\Delta m > 100$~GeV region
as can be found in figure~\ref{fig:efficiencies}.
The MC errors are less than $10\%$ for all $\Delta m$. The efficiencies remain valid also for the case of three-body decays.
The method described above allows us to assess the exclusion in a first approximation for similar SUSY models without 
carrying out a detailed MC simulation.

\begin{table}[t]
\renewcommand{\arraystretch}{1.4}
\begin{center}
\begin{tabular}{c|c||c|c|c|c|c|c|c} 
\toprule
\multicolumn{2}{c||}{$\Delta m$ (GeV)}                        & 50 & 60 & 70 & 80 & 100 & 120 & 140 \\
\hline
         & $\tilde \chi_1^{\pm} \tilde \chi_1^{\mp}$  & $0.21$ & $ 0.53$ & $ 0.87$ & $1.30$ & $1.9$ & $ 2.6$ & $3.0$    \\
2OS  & $\tilde \chi_1^{\pm} \tilde \chi_2^0$                  & $0.18$ & $0.41$ & $0.61$ & $0.82$ & $1.07$ & $1.29$  & $1.34$ \\
          & $\tilde \chi_2^0 \tilde \chi_2^0$               & $0.18$ & $0.27$ & $0.37$ & $0.42$ & $0.52$ & $0.53$  & $0.48$ \\          
          & $\tilde \chi_2^0 \tilde \chi_1^0$          & $0.26$ & $0.70$ & $1.31$ & $1.85$ & $3.52$ & $4.43$ & $4.33$ \\  
\hline
         & $\tilde \chi_1^{\pm} \tilde \chi_1^{\pm}$  & $0.79$ & $2.23$ & $4.40$ & $7.00$ & $12.93$ & $18.24$ & $22.34$    \\
2SS  & $\tilde \chi_1^{\pm} \tilde \chi_2^0$                  & $0.38$ & $0.88$ & $1.46$ & $2.00$ & $3.16$ & $4.22$ & $5.25$    \\
          & $\tilde \chi_2^0 \tilde \chi_2^0$               & $0.36$ & $0.70$ & $1.03$ & $1.28$ & $1.28$ & $2.12$ & $2.08$    \\
\hline
            & $\tilde \chi_1^{\pm} \tilde \chi_1^{\pm}$  & $0.55$ & $1.54$ & $2.79$ & $4.63$ & $7.76$ & $9.23$ & $11.70$    \\
2SS+  & $\tilde \chi_1^{\pm} \tilde \chi_2^0$                  & $0.25$ & $0.56$ & $0.89$ & $1.18$ & $1.82$ & $2.11$ & $2.80$    \\
            & $\tilde \chi_2^0 \tilde \chi_2^0$               & $0.25$ & $0.46$ & $0.63$ & $0.78$ & $0.63$ & $1.03$ & $1.00$    \\
\hline
3LEP  & $\tilde \chi_1^{\pm} \tilde \chi_2^0$                  & $0.16$ & $0.73$ & $2.16$ & $4.41$ & $10.39$ & $14.75$ & $17.33$  \\
            & $\tilde \chi_2^0 \tilde \chi_2^0$               & $0.43$ & $1.58$ & $4.01$ & $7.70$ & $12.35$ & $15.91$ & $18.03$    \\
\hline
3LEP+  & $\tilde \chi_1^{\pm} \tilde \chi_2^0$                  & $0.15$ & $0.71$ & $2.05$ & $4.07$ & $8.57$ & $10.17$ & $8.83$ \\
              & $\tilde \chi_2^0 \tilde \chi_2^0$               & $0.42$ & $1.55$ & $3.87$ & $7.01$ & $10.18$ & $10.53$ & $8.23$  \\
\hline
4LEP    & $\tilde \chi_2^0 \tilde \chi_2^0$               & $0.28$ & $1.24$ & $3.73$ & $8.50$ & $15.76$ & $18.83$ & $23.61$    \\            
\bottomrule
\end{tabular}
\end{center}
\renewcommand{\arraystretch}{1.0}
\caption{Comparison of efficiency in $\%$ for different signal regions and intermediate states as a function of mass difference, $\Delta m
= m_{\tilde{g}} - m_\mathrm{LSP}$. Gluino mass was set to $m_{\tilde{g}} = 800$~GeV.
 \label{tab:eff}}
\end{table}

\section{Summary and conclusions \label{summary}}

We have studied the impact of ATLAS multi-lepton searches on compressed SUSY models at $\sqrt{s} = 7$~TeV. We introduced a class of compressed models
with leptonic gaugino decays to study the discovery reach of the analyses. We parametrise our simplified models by two key parameters, the gluino mass and 
the mass difference between gluino and the LSP, $\Delta m$.
In order to improve discovery potential we modified some of the ATLAS searches
by imposing additional cuts motivated by the properties of compressed models.  
The best exclusion is obtained for a tri-lepton search with an additional cut on the leading lepton, $p_T < 40 \gev$. 
Gluino mass up to 950~GeV can be excluded for a mass difference between gluino and the LSP as small as 100~GeV. Further improvement can be expected when the full 2011 data sample will be included in the analysis. These limits can compete 
with standard jets+\met\ searches. 

We also discuss in detail the efficiency of event selection in each of the studied signal regions. Decreasing the mass difference between gluino and the LSP leads to a substantial decrease in selection efficiency, that varies as well with gluino mass. This arises due to lepton $p_T$ cuts and a cut on invariant mass of SFOS lepton pairs. We analyse the topologies classified according to different intermediate states and calculate efficiency for each of them separately. This allows for a comparison with different models and, in principle, quick assessment of exclusion limits for models with different branching ratios and cross sections without carrying out MC simulation.  

We also studied some of the properties of the compressed SUSY models. We point out that lepton pair invariant masses and jet $p_T$ distribution 
could give insight into the intermediate mass scales in the model. In particular, jet $p_T$ distribution can provide an information on the separation of the SUSY-QCD and gaugino mass scale.

In conclusion, lepton searches can provide interesting limits on compressed supersymmetric models, which could help to overcome limitations of usual jet plus \met\ search channels and further constrain available parameter space. A similar analysis using CMS data may strengthen the exclusion limits as the CMS experiment is using different lepton selection criteria. This problem we leave for a forthcoming study.

\acknowledgments
We would like to thank Simon Pl\"{a}tzer for useful discussions.

\bibliographystyle{JHEP}
\bibliography{compressed}

\end{document}